\documentclass[preprint,showpacs,preprintnumbers,amsmath,amssymb]{revtex4}

\usepackage{epsfig,graphicx}
\usepackage{dcolumn}
\usepackage{bm}

\begin{document}
\title{Simulation and theory of vibrational phase relaxation in the critical and
supercritical nitrogen: Origin of observed anomalies}

\author{Swapan Roychowdhury\footnote{Author to whom correspondence
should be address; electronic mail: swapan@sscu.iisc.ernet.in}
 and Biman Bagchi}

\affiliation{Solid State and Structural Chemistry Unit,
        Indian Institute of Science.
        Bangalore 560 012,
        India}

\begin{abstract}
 We present results of extensive computer simulations and theoretical analysis 
of vibrational phase relaxation of a nitrogen molecule {\it along the critical 
isochore} and also along the gas-liquid coexistence. The simulation includes 
all the different contributions [atom-atom (AA), vibration-rotation (VR) and 
resonant transfer] and their cross-correlations. Following Everitt and Skinner, 
we have included the vibrational coordinate ($q$) dependence of the interatomic 
potential. It is found that the latter makes an important contribution. The 
simulated results are in good agreement with the experiments. Dephasing time 
($\tau_{\it v}$) and the root mean square frequency fluctuation ($\Delta$) in 
the supercritical region are calculated. The principal important results are: 
(a) a crossover from a Lorentzian-type to a Gaussian line shape is observed as 
the critical point is approached along the isochore (from above), (b) the root 
mean square frequency fluctuation shows nonmonotonic dependence on the 
temperature along critical isochore, (c) along the coexistence line and the 
critical isochore the temperature dependent linewidth shows a divergence-like 
$\lambda$-shape behavior, and (d) the value of the critical exponents along the 
coexistence and along the isochore are obtained by fitting. It is found that 
the linewidths (directly proportional to the rate of vibrational phase 
relaxation) calculated from the time integral of the normal coordinate time 
correlation function [$C_{Q}(t)$] are in good agreement with the known 
experimental results. The origin of the anomalous temperature dependence of 
linewidth can be traced to simultaneous occurrence of several factors, (i) 
the enhancement of negative cross-correlations between AA and VR contributions 
and (ii) the large density fluctuations as the critical point (CP) is 
approached. The former makes the decay faster so that {\it local density 
fluctuations} are probed on a femtosecond time scale. The reason for the negative 
cross-correlation between AA and VR is explored in detail. A mode coupling 
theory (MCT) analysis shows the slow decay of the enhanced density 
fluctuations near critical point. The MCT analysis demonstrates that the large 
enhancement of VR coupling near CP arises from the non-Gaussian behavior of 
density fluctuation and this enters through a nonzero value of the triplet 
direct correlation function.
\end{abstract}

\maketitle

\section{Introduction}
 The study of vibrational phase relaxation (VPR) has been an important endeavor of 
a physical chemist/chemical physicist in the attempt to understand and quantify 
the interaction of a chemical bond with the surrounding solvent molecules 
 \cite{oxrev1,oxsim}. As the phase relaxation of a bond is sensitive to the 
details of intermolecular interaction, it is a useful tool in understanding 
various aspects of solute-solvent interactions. The study of VPR has been driven 
by the sensitive experimental measurements of the vibrational linewidth. Thus the 
theoretical results can be verified with experiments directly. Simple models 
such as the isolated binary collisions model \cite{fish} and hydrodynamic model 
\cite{oxrev1} were employed initially to explain the experimental results. These 
models led to simple expressions of temperature, density, and viscosity dependence 
of the rate. However, the agreement with experiments was at most tentative. It 
was soon realized that a major difficulty was that dephasing derives 
contributions from many sources and there may not be any unique mechanism of 
dephasing. Not only the estimation of these contributions are nontrivial, 
even the cross-correlations between different pure terms are non-negligible. 
Thus the Kubo theory of dephasing \cite{kubo} was straightforward to apply in 
this case and the calculations of the frequency modulation time correlation 
function turned out to be extremely difficult, even for simple diatoms like 
nitrogen (N$_{2}$).

    Uniqueness of vibrational dephasing near the critical point is that the
value of mean square frequency fluctuation $\large<\Delta\omega^{2}(0)\large>$ 
becomes large, leading to rapid decay of $\large<Q(t)Q(0)\large>$. Many factors 
which are responsible for this behavior are very difficult to understand. Near 
high temperature vibration-rotational (VR) coupling shows large enhancement to 
$\large<\Delta\omega(t)\Delta\omega(0)\large>$ including negative 
cross-correlation between atom-atom and VR coupling terms.
 
    Many experimental studies have been carried out on N$_{2}$ using vibrational 
Raman spectroscopy as a probe. Experimental studies of Clouter {\it et al.} 
\cite{clouter1,clouter2} showed that the isotropic Raman line shape of simple 
fluid like $N_{2}$ may exhibit a remarkable additional broadening near liquid-gas 
critical points ($\rho_{c},T_{c}$). They measured the Raman spectra along 
the triple point to the critical point and behavior of the line shape as the 
critical point is approached from above at a constant density.
Recently Musso {\it et al.} \cite{musso} calculated 
the important cross-correlations between resonant and nonresonant dephasing 
mechanisms in dense liquid. They observed an interesting temperature 
dependence $\lambda$ shaped linewidth ($\Gamma$) along the coexistence and 
along the critical isochore. 
        
     In their pioneering study, Oxtoby {\it et al.} \cite{oxrev1,oxsim,oxrev2} 
showed that direct simulation of the vibrating molecules could be avoided for 
most cases of interest. A quantum mechanical perturbation theory for the 
vibrational motion can be used to express the dephasing rate in terms 
of auto- and cross-correlation functions of bond-force terms and its 
derivatives. The latter ones can then be calculated by molecular dynamics (MD) simulations. 
Oxtoby {\it et al.} calculated the linewidth and the motionally narrowed line 
shape of nitrogen near boiling point (77 K) \cite{fish,oxrev2}. They considered 
several contributions from (i) solvent-solute interaction force in the liquid, 
(ii) their derivatives, and (iii) resonant molecular vibrational interactions to 
the frequency fluctuation of molecules.

   Recently Gayathri {\it et al.} \cite{gay1,gay3} calculated the vibrational 
phase relaxation of the fundamental and the overtones \cite{tom1,tom2} of the N-N 
stretch of nitrogen in pure nitrogen by molecular dynamics (MD) simulations. They 
reproduced the experimental data semiquantitatively (within 40\% in most cases). 
They have also applied the mode coupling theory (MCT) \cite{sarika1,
swapan} to compare with the simulation results. In their calculations they have not 
include the vibrational coordinate ($q$) dependence of the interatomic potential 
and also ignored the cross-terms among the vibration-rotation coupling term, the 
atom-atom term, and the resonance term. More recently Everitt and Skinner studied 
the Raman line shape of nitrogen in a systematic way by including the bond length 
dependence of the dispersion and repulsive force parameters \cite{eve}. They have 
also included the cross-correlation terms which were neglected earlier and the 
results for the line shift and the linewidth along the liquid-gas coexistence of 
N$_{2}$ were observed to be in very good agreement with experiments. But calculations 
along the critical isochore have not been reported in their study. As mentioned 
earlier, there exists a profound experimental results in this region. 

          In this work, we report results of extensive MD simulations of 
vibrational dephasing along critical isochore. We have calculated the linewidth, 
the line shape, and the dephasing time of N$_{2}$ {\it along the critical isochore} 
and along the coexistence line. The normal coordinate time correlation function 
[$C_{Q}(t)$] is calculated from the frequency fluctuation time correlation function 
for different state points along the coexistence line as well as along the 
isochore of N$_{2}$. 

 We have incorporated the vibrational coordinate ($q$) dependency of the 
intermolecular potential and also the cross-terms. The linear expansion of the 
Lennard-Jones potential parameters on vibrational coordinate is very important 
to get the correct sign of line shift. The time integral of the diagonal and cross-terms 
of frequency fluctuation time correlation function [$C_{\omega}(t)$] gives the 
contribution to the linewidth. These cross-terms have a large effect on the 
linewidth to get a good agreement with experiment.   
  
     The line shape calculated from the normal coordinate time correlation function 
shows the Gaussian behavior close to the critical point. Experimentally 
\cite{musso}, it has been proved that the line shape remains Lorentzian for the 
liquid near its normal boiling point (BP). The increase in density fluctuations 
near the critical point increases the mean square frequency fluctuation 
$\large<\Delta\omega^{2}_{i}\large>$, transferring the line shape from its fast modulation, 
i.e., Lorentzian shape limit outside the critical region to a slower modulated 
Gaussian shape. The root mean square (rms) frequency fluctuation of N$_{2}$ 
calculated along the isochore shows nonmonotonic behavior. However, the 
dephasing time ($\tau_{v}$) did not show any nonmonotonicity. 

\section{Basic Expressions }

  The theories of the vibrational dephasing are all based on Kubo's stochastic 
theory of the line shape. This gives a simple expression for the isotropic Raman 
line shape $[\it I(\omega)]$ in terms of Fourier transform of the normal 
coordinate time correlation function [$C_{Q}(t)$] through the polarizability 
time-correlation function as given by \cite{kubo,kubo1}, 
\begin{eqnarray}
I(\omega)=\int_{0}^{\infty}dt\exp(i\omega t)\left[\large<Q(t)Q(0)\large>\right].
\label{eq:isoline}
\end{eqnarray}
  A cumulant expansion of Eq.~(\ref{eq:isoline}) followed by truncation after 
second order gives the following well known expression for $C_{Q}(t)$
(Ref~\cite{kubo}):
\begin{eqnarray}
C_{Q}(t)&=&\large<Q(t)Q(0)\large>\nonumber\\&=&Re \exp(i\omega_{0}t+i\large<\Delta\omega\large>t)
\nonumber\\&\times&\exp\left[-\int_{0}^{t}dt^{\prime}(t-t^{\prime})\large<\Delta
\omega(t^{\prime})\Delta\omega(0)\large>\right],
\label{eq:cqt}
\end{eqnarray}
where $\Delta\omega_{i}(t) = \omega_{i}(t) - \large<\omega_{i}\large>$ is the fluctuation 
of the vibrational frequency from average vibrational frequency. 
$\large<\Delta\omega(t)\Delta\omega(0)\large>$ is the frequency fluctuation time 
correlation [$C_{\omega}(t)$] function and $\omega_{0}$ is the fundamental 
vibrational frequency of nitrogen. 

  The fluctuation in energy between the ground state and the $n$th quantum level 
of overtone transitions is given by
\begin{equation}
\hbar\Delta\omega_{n0}^{i}(t) = V_{nn}^{i}(t)-V_{00}^{i}(0)+ 
\sum_{j}V_{ij}(t)~.
\label{eq:delomegait}
\end{equation}
$V_{nn}^{i}$ is the Hamiltonian matrix element of the coupling of the vibrational 
mode to the solvent bath and $\sum_{j}V_{ij}(t)$ represents the contribution from 
resonant energy transfer between two molecules $i$ and $j$. 

 The Hamiltonian for the normal mode ($Q$) is assumed to be of the following 
anharmonic form:  
\begin{equation}
{\it H}_{vib} = \frac{1}{2}\mu\omega_{0}^{2}Q^{2} + \frac{1}{6}
{\it f}Q^{3},
\label{eq:vibham}
\end{equation}
where $\mu$ is the reduced mass and $f$ is the anharmonic force constant. 
The value of $f$ is $17.8\times 10^{4}$ gm/cm s$^{2}$. Note that $Q$ in 
Eq.~(\ref{eq:vibham}) is not in the mass-weighted form.

 If $V$ is the oscillator-medium interaction potential, then one finds the 
following expression for the fluctuation in overtone frequency (by using 
perturbation theory) 
$\Delta\omega_{n0}(t)$ Ref~(\cite{oxsim}):
\begin{widetext}
\begin{eqnarray}
\hbar\Delta\omega^{i}_{n0}(t)&=&(Q_{nn} - Q_{00})\left(\frac{\partial V} 
{\partial Q}\right)_{Q=0}(t) + \frac{1}{2}[(Q^{2})_{nn} 
- (Q^{2})_{00}]\left(\frac{\partial^{2}V}{\partial^{2}Q}\right)_{Q=0}(t)  
\nonumber \\&+& Q^{2}_{n0}\sum_{j\ne i}\left(\frac{\partial^{2}V}
{\partial Q_{i}Q_{j}}\right)_{Q=0}(t) + \ldots \nonumber \\
&=&\left(\frac{{\it n}\hbar(-{\it f})}{2\mu^{2}\omega^{3}_{0}}\right)
{\it F}^{i}_{1Q} + \left(\frac{{\it n}\hbar}{2\mu\omega_{0}}\right)
{\it F}^{i}_{2Q} + \delta_{n1}\left(\frac{\hbar}{2\mu\omega_{0}}
\right)^{\frac{1}{2}}\sum_{i \ne j}{\it F}^{\it i\it j}_{3Q}~.
\label{eq:delomeganit}
\end{eqnarray}
\end{widetext}
The first two terms in the right-hand side of Eq.~(\ref{eq:delomeganit}) are 
the atom-atom contributions and the third term is the resonance term. 

  The vibration-rotation (VR) contribution to the broadening of the line shape 
is given by $\Delta\omega_{n0,VR} = \Delta R_{n0}/\hbar{\it I}_{m}
{\it r}_{e}. \Delta {\it J}^{2}$ \cite{schandler,brueck}; where $\Delta
{\it J}^{2}(t)={\it J}^{2}(t) - \large<{\it J}^{2}(0)\large>$. However, the time correlation 
function of VR coupling term is given by
\begin{eqnarray}
C^{VR}_{\omega}(t)&=&\large<\Delta\omega(t)\Delta\omega(0)\large>_{n0,VR}\nonumber \\
&=&\left(\frac{\Delta R_{n0}}{\hbar{\it I}_{m}{\it r}_{e}}\right)^{2}
\times(\large<{\it J}^{2}(t){\it J}^{2}(0)\large>
\nonumber\\&&-\large<{\it J}^{2}\large>^{2})~.
\label{eq:omegavr1}
\end{eqnarray}
${\it J}$ is the angular momentum and ${\it I}_{m}$ is the moment of inertia 
value at the equilibrium bond length(${\it r}_{e}$).  
  
  Auto- and cross-correlations between atom-atom forces, VR coupling and resonance 
terms have been considered in our model. The final expression for $C^{total}_{
\omega}(t)$ can thus be written as
 \begin{eqnarray}
C^{total}_{\omega}(t)&=&\large<\Delta\omega(t)\Delta\omega(0)\large>\nonumber\\
&=&<\Delta\omega_{AA}(t)\Delta\omega_{AA}(0)\large>\nonumber\\ 
&+& \large<\Delta\omega_{VR}(t)\Delta\omega_{VR}(0)\large>\nonumber\\
&+& \large<\Delta\omega_{Rs}(t)\Delta\omega_{Rs}(0)\large>\nonumber\\
&+& \large<\Delta\omega_{AA}(t)\Delta\omega_{VR}(0)\large>\nonumber\\
&+& \large<\Delta\omega_{VR}(t)\Delta\omega_{Rs}(0)\large>\nonumber\\
&+& \large<\Delta\omega_{Rs}(t)\Delta\omega_{AA}(0)\large>~.
\label{eq:omegatotal}
\end{eqnarray}

  The line shape is obtained from the frequency fluctuation correlation function. 
${\it F}^{i}_{1Q},{\it F}^{i}_{2Q}$ and ${\it F}^{i}_{3Q}$ in 
Eq.~(\ref{eq:delomeganit}) and the 
VR coupling from Eq.~(\ref{eq:omegavr1}) are calculated separately. The main 
difference from a previous calculation \cite{gay4} is that all the terms are 
vibrational coordinate dependent. This dependency comes through the bond 
length of the ({\bf r} = {\bf r}($q$)) molecule. 

   The Hamiltonian of homonuclear diatomic molecules can be expressed as the
sum of three terms,
\begin{equation}
{\rm H} = {\rm H}_{\it v} + {\rm T}(\bf q) + {\rm U}(\bf q),
\label{eq:totalham}
\end{equation}
${\rm H}_{\it v}$ is the vibrational Hamiltonian. ${\rm T}(\bf q)$
is the total translational and rotational kinetic energy. ${\rm U}(\bf q)$ 
is the inter-molecular potential energy. ${\bf q}$ represents the collection 
of vibration coordinates $\{q_{\it i}\}$. The vibration Hamiltonian for
the isolated (gas-phase) molecules is given by 
\begin{equation}
{\rm H}_{\it v} = \sum_{\it i} \left(\frac{{\it p}^{\rm 2}_{\it i}}{2\mu} +
{\it u}(q_{\it i})\right).
\end{equation}
Here $\sum$ represents the sum of all anharmonic oscillators for the vibrational 
modes of gas molecules. The conjugate momentum of $q_{\it i}$ is $p_{\it i}$ for 
the ${\it i}$th molecule of the oscillator. The translational and rotational 
kinetic energy term of the oscillator can be written as
\begin{equation}
{\rm\vec T}(\bf q) = \sum_{\it i} \left(\frac{{\it\bf P}^{\rm 2}_{\it i}}{2M} +
\frac{{\it\bf L}^{\rm 2}_{\it i}}{2{\rm I}(q_{\it i})}\right),
\end{equation}
where ${\rm\bf P_{\it i}}$ and ${\rm\bf L_{\it i}}$ are the 
center of mass momentum and the angular momentum for molecule ${\it i}$,
respectively and ${\rm I}(q) = \mu ({\rm r}_{\it e} + q)^{2}$
is the moments of inertia.
We can express ${\rm H} = {\rm T}(0) + {\rm U}(0)$ as a bath Hamiltonian 
and the perturbation Hamiltonian {\rm V} is given by, 
${\rm V} = {\rm T}(\bf q)+ {\rm U}(\bf q)-{\rm T}(0) - {\rm U}(0)$.
The total Hamiltonian 

\begin{equation}
{\rm H} = {\rm H}_{\it v} +{\rm H}_{\it b} + {\rm V}. 
\end{equation}

\begin{figure}
\epsfig{file=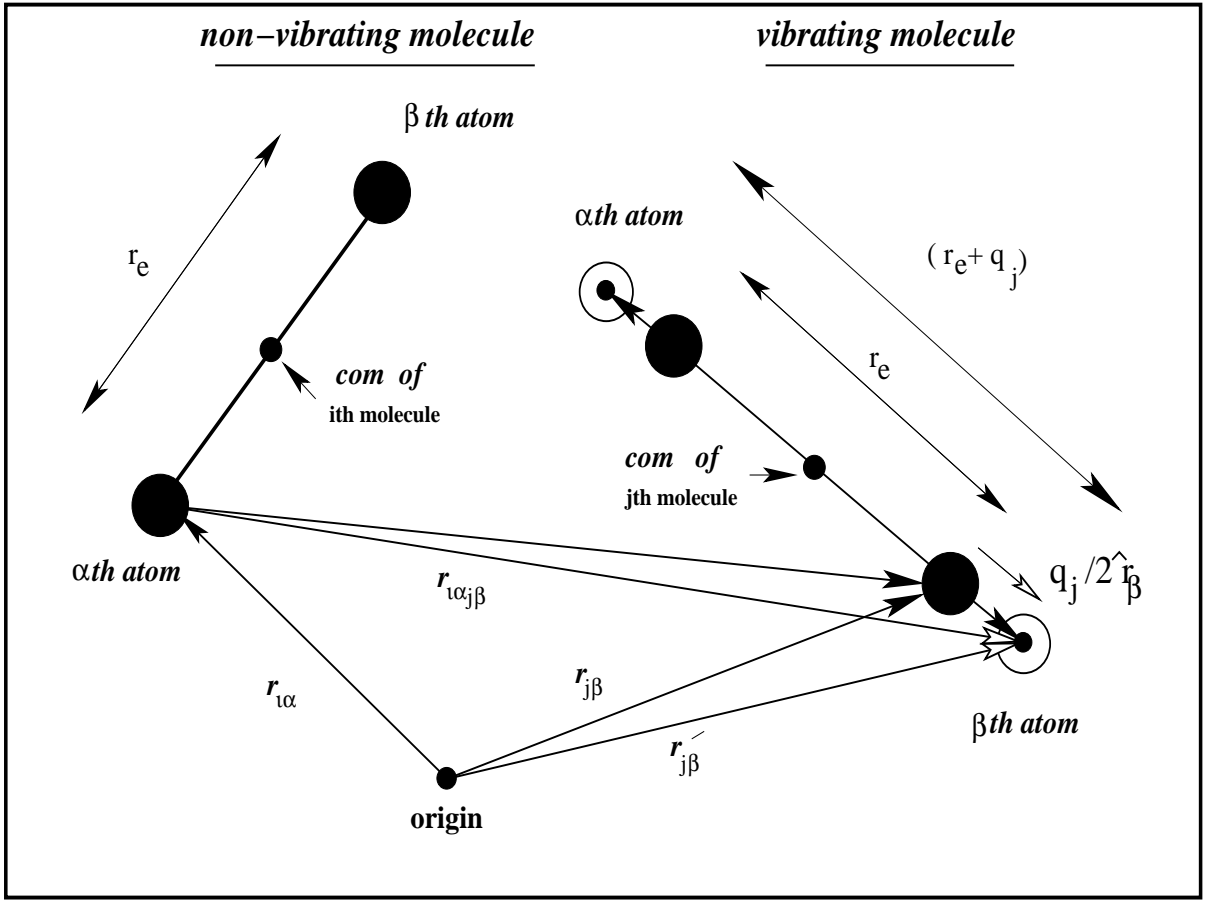,height=12cm,width=16cm}
\caption{\label{fig:FIG1}The schematic diagram illustrating  molecular 
interactions between the nonvibrating ($i$th molecule) and vibrating 
($j$th molecule) homonuclear diatomic (N$_{2}$).}
\end{figure}

    The intermolecular potential energy can be written as 
\begin{equation}
{\rm U}(\bf{\rm q}) = \frac{1}{2}\sum_{{\it i}\neq{\it j}}\sum_{\alpha\beta} 
{\it v}\\ (\epsilon_{\it ij},\sigma_{\it ij}, {\bf {\rm r}}_{{\it i}\alpha
{\it j}\beta}),
\end{equation}
with
\begin{equation}
{\bf r}_{i\alpha j\beta} = {\bf {r^\prime}}_{j\beta} - {\bf r}_{i\alpha} \\
= {\bf r}_{j\beta} + \frac{q}{2}{\hat{\rm r}}_{j\beta} -{\bf r}_{i\alpha}, 
\end{equation}
and
\begin{eqnarray}
{\vec {r^\prime}}_{j\beta}(\vec q_{\it j}) = {\vec r}_{j\beta}(0)
+ \frac{q}{2}{\hat{\rm r}}_{j\beta}. 
\end{eqnarray}
Here ${\hat{\bf r}}_{j\beta}$ is the unit vector along $\beta$ atom of the vibrating 
molecule ($i$th molecule) from the center of mass (see Fig.~$\ref{fig:FIG1}$).

\section{Simulation details}

\begin{figure}
\epsfig{file=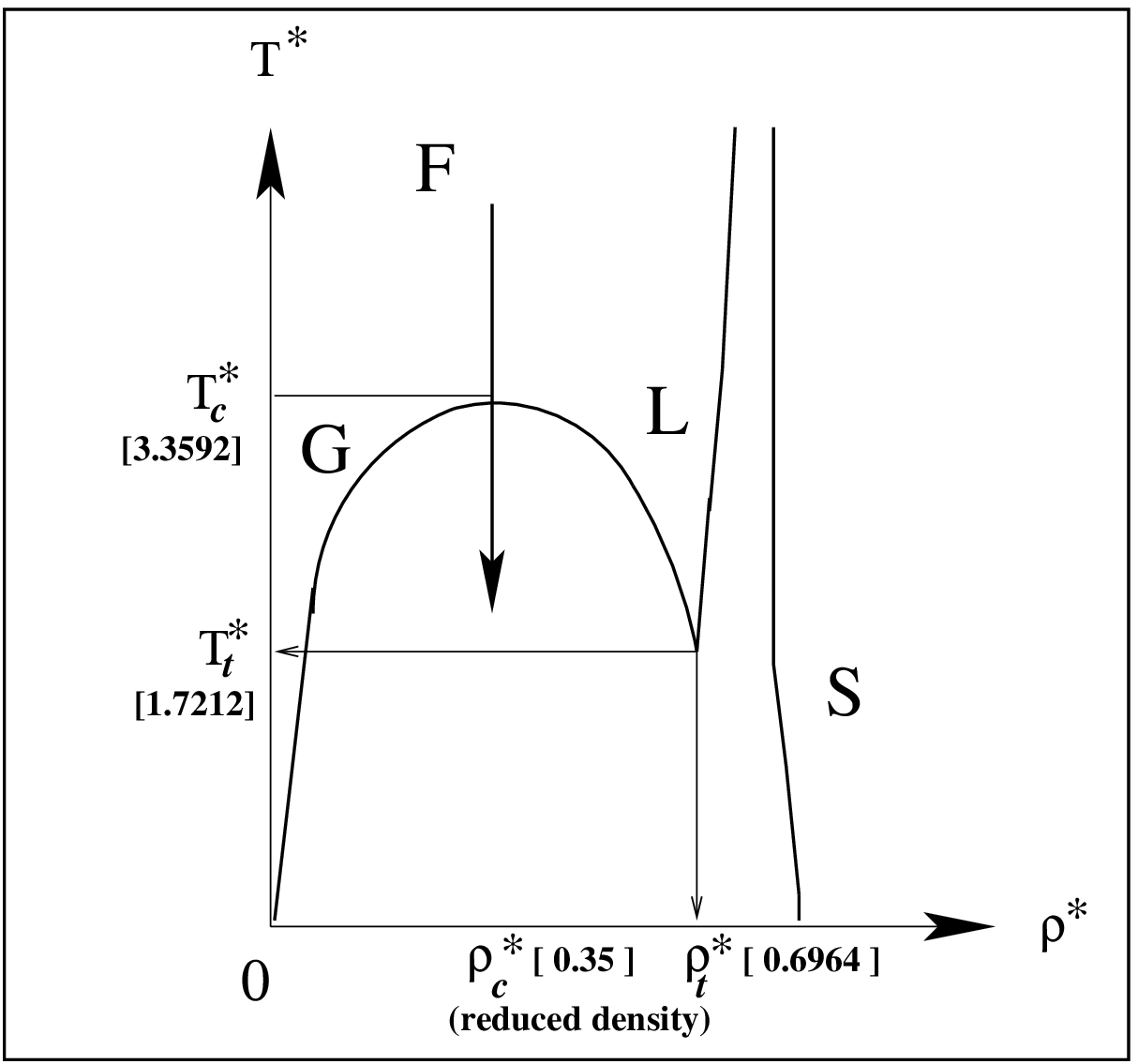,height=12cm,width=14cm}
\caption{\label{fig:FIG2}Phase diagram of a typical substance, showing 
boundaries between solid (S), liquid (L) and vapor (G) or fluid (F) phases. This 
is only the projection of the $\rho^{*}-T^{*}$ plane. The parameters ($T^{*}_{c}$, 
$T^{*}_{t}$, $\rho^{*}_{c}$, and $\rho^{*}_{t}$) are given for nitrogen. We did 
the simulation along the critical isochore of nitrogen which is indicated 
by an arrow.}
\end{figure}

  We have performed a microcanonical ({\it NVE}) ensemble molecular-dynamics (MD) 
simulation \cite{allen,graphite} at different state points of N$_2$ ranging from 
the melting point (also the triple point of N$_2$) through the boiling point and 
along the critical isochore (see Fig.~$\ref{fig:FIG2}$) using the leap-frog 
algorithm \cite{fincham}. The parameters used are given in ~\ref{tab:table1}
\cite{herz}. A system of 256 $N_{2}$ diatomic molecules were enclosed in a cubic 
box and periodic boundary conditions were used.

\begin{table}
\caption{\label{tab:table1}Parameters for $N_{2}$.}
\begin{ruledtabular}
\begin{tabular}{lcr}
Potential parameters & $r_{o}/\AA$ & 1.094\\
 & $\epsilon/kK$  &  37.3 \\   
 & $\sigma /\AA$ & 3.31 \\\hline
Spectroscopic Constants & M/amu  & 28.0\\
 & $\omega_{o}/cm^{-1}$ & 2358.57 \\\hline
Polarizability parameters & $\gamma /\AA^{-1}$ & 0.62\\
& $\delta /\AA^{-1}$ &  -0.063 \\ 
\end{tabular}
\end{ruledtabular}
\end{table}

   The thermodynamic state of the system is expressed in terms of the reduced 
number density of $\rho^{*} = \rho\sigma^{3}$ (Ref.~\cite{allen}) and a reduced 
temperature of $T^{*} = k_{B}T/\epsilon$. $\sigma$ is the diameter of the 
molecule and $\epsilon$ is the interaction parameter (see~\ref{tab:table1}). The 
unit of temperature is $\epsilon/k_{B}$ (K), where $k_{B}$ is the Boltzmann 
constant. Cheung and Powels \cite{cheung} had earlier studied liquid N$_{2}$ 
at different state points using MD simulations. Most of the thermodynamic 
state points chosen for the work presented here have been taken from their 
study. We have done few simulations with a system of 512 nitrogen molecules 
to check the system size dependency.

  Fig.~$\ref{fig:FIG2}$ gives a schematic view of the phase diagram 
\cite{balzop}. The arrowed line points out that along the critical isochore, 
$T_{c}$ is approached from above. For nitrogen the triple point corresponds 
to that given by ($T^{*}_{t},\rho^{*}_{t})=(1.7212, 0.6964$) and the critical point, 
$(T^{*}_{c},\rho^{*}_{c}) =(3.3592, 0.35)$.

 For intermolecular potential-energy (${\it V}_{\it ij}$) between two molecules 
{\it i} and {\it j}, the following site-site Lennard-Jones type is employed as 
given below, 
\begin{equation}
{\rm V}_{\it ij}=\sum_{\alpha,\beta}^{\it 1,2}{\rm V}({\rm r}_{i\alpha j\beta}).
\end{equation}
Here ${\rm V}({\rm r}_{i\alpha j\beta})$ is the Lennard-Jones atom-atom 
potential defined as
\begin{equation}
{\rm V}({\rm r}_{i\alpha j\beta})= 4{\epsilon}_{i\alpha j\beta}[(\frac{{\it
\sigma}_{i\alpha j\beta}}{{\rm r}_{i\alpha j\beta}})^{12}-(\frac{{\sigma}_{i
\alpha j\beta}}{{\rm r}_{i\alpha j\beta}})^6].
\end{equation} 
Vibrational coordinate dependence of $\epsilon$ and $\sigma$ has been 
incorporated following Everitt and Skinner \cite{eve},
\begin{eqnarray}
\epsilon(q) &=& \epsilon[1 + 2\gamma q];\nonumber\\\sigma(q) &=& 
\sigma[1+ 2\delta q]. 
\end{eqnarray}
For homonuclear diatomic-like nitrogen,
\begin{eqnarray}
\epsilon_{i\alpha j\beta} &=& \sqrt{\epsilon_{i\alpha}\epsilon_{j\beta}}
\nonumber \\&=&\epsilon_{ij},
\end{eqnarray}
\begin{eqnarray}
\sigma_{i\alpha j\beta} &=& \frac{\sigma_{\alpha\alpha}+\sigma_{\beta\beta}}{2}
\nonumber\\ &=&\sigma_{ij}, 
\end{eqnarray}
and 
\begin{equation}
\epsilon_{ij}\simeq \epsilon(1+\gamma q_{i}+\gamma q_{j} + 2\gamma^{2}
q_{i}q_{j}),
\end{equation}
\begin{equation}
\sigma_{ij} = \sigma(1+\delta q_{i}+\delta q_{j}).
\end{equation} 
Now the LJ potential takes the form as below,  
\begin{eqnarray}
{\rm V}_{ij} &=& \sum_{\alpha,\beta = 1}^{2}\large[4\epsilon \{1+\gamma(q_{i}
+ q_{j})+ 2\gamma ^{2}q_{i}q_{j}\}\nonumber\\&&\times\large\{\large(\frac{\sigma}
{\rm r_{i\alpha j\beta}}\large)^{12}(1+\delta q_{i} + \delta q_{j})^{12}\nonumber \\
&&-\large(\frac{\sigma}{\rm r_{i\alpha j\beta}}\large)^{6}(1+
\delta q_{i} + \delta q_{j})^{6}\large\}\large]. 
\end{eqnarray}
\section{Simulation Results, Comparison with Experiment and  
Discussion}
\subsection{Along the coexistence line}
       The frequency-modulation time correlation function {\large[$C_{\omega}(t)$\large]}, the 
dephasing linewidth, and the line shift \cite{amot} are all obtained for several 
thermodynamic state points ranging from triple point to critical point. 

\begin{figure}
\epsfig{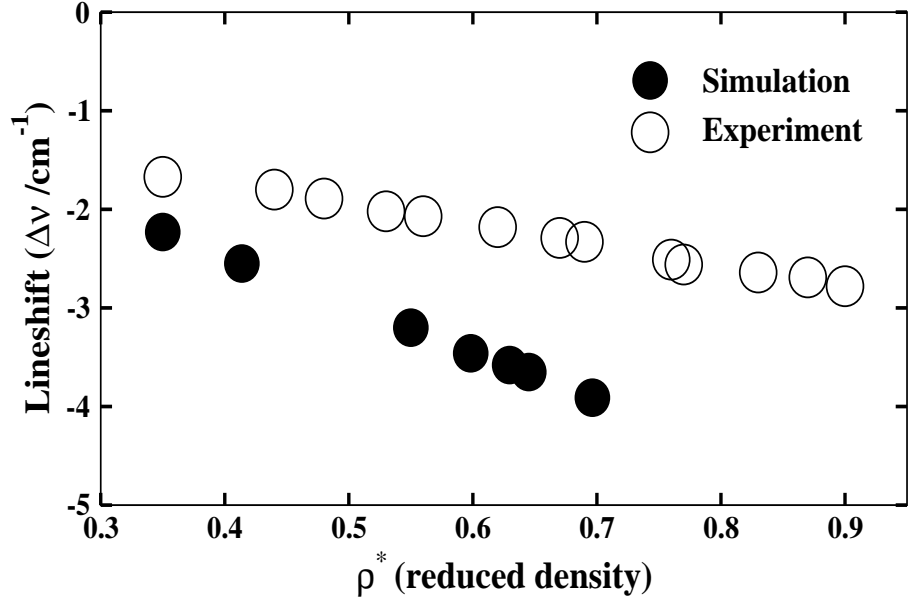}\\
\vspace*{3cm}
\epsfig{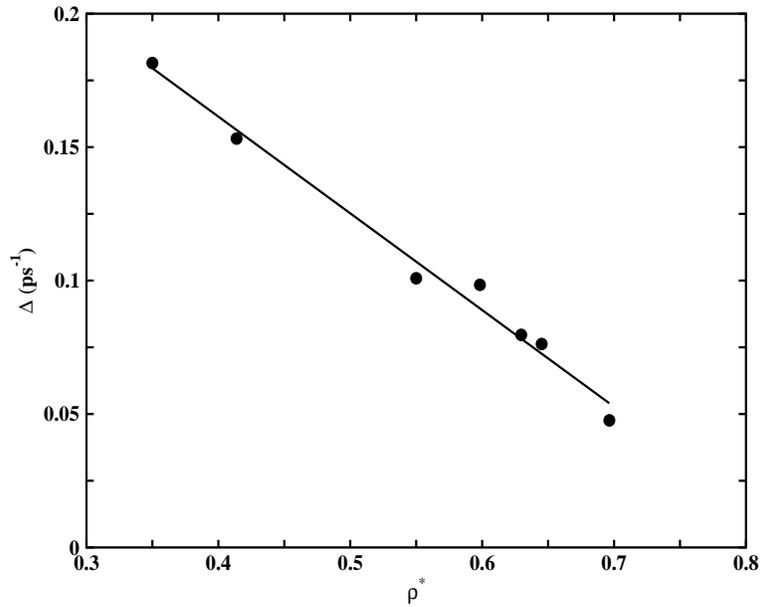}
\caption{\label{fig:FIG3}The Raman line shift ($\Delta\nu$) is plotted against density 
($\rho^{*}$) ($a$) along the coexistence line for N$_{2}$. The critical density is 
indicated by the arrow. The rms frequency fluctuation $\Delta$ is plotted against
reduced density ($b$) along the coexistence line for N$_{2}$. Simulation 
results have been fitted with the formula $y = ax + b$ with $ a = - 0.3624$ and 
$b = 0.3064$.}
\end{figure}

      We have calculated the line shift $\Delta\nu^{~} = (\left<\omega_{i}\right>
- \omega_{0})/2\pi c$ as a function of density for nitrogen. The magnitude of the
line shift increases with density as clearly seen in Fig.~$\ref{fig:FIG3}$(a) and the results are 
in good agreement with the experiment. The vibrational coordinate dependence is 
important to get the correct sign of the line shift.

   Along the coexistence line the root mean square frequency fluctuation ($\Delta$) 
increases as the critical point is approached. Fig.~$\ref{fig:FIG3}$(b) shows that 
$\Delta$ decreases linearly with density with a slope of -0.3624.

\subsection{Along the critical isochore}

\begin{figure}
\epsfig{file=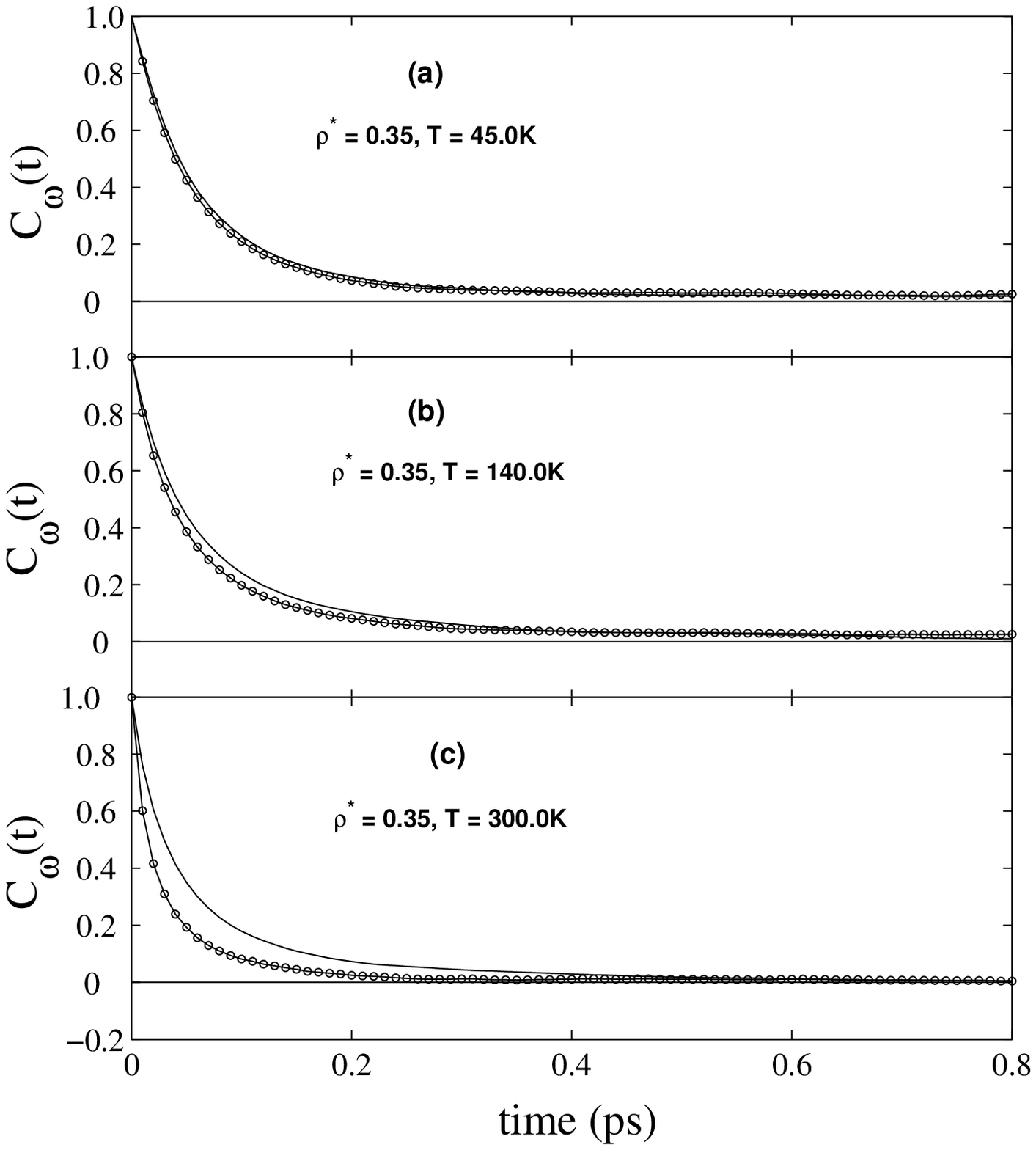}
\caption{\label{fig:FIG4}The frequency fluctuation time correlation functions 
{\Large[$C_{\omega}(t)$\Large]} are plotted against time (t) at different temperatures (a) 
T = 45.0 K, (b) T = 140.0 K, and (c) T = 300.0 K along the critical isochore. In all the 
figures above, circles with a simple line show $C_{\omega}(t)$ for $q$ dependent 
interaction potential, whereas the simple solid line shows the $q$ 
independent interaction potential.}
\end{figure}

  We have plotted  $C_{\omega}(t)$ against time for three different temperatures 
along the critical isochore in Fig.~$\ref{fig:FIG4}$. The lines with open circles show 
$C_{\omega}(t)$ which decays faster than $C_{\omega}(t)$ represented by the 
simple line.

      The temperature dependence of the dephasing time ($\tau_{v}$) calculated by 
integration of $C_{\omega}(t)$ {\large[Eq.~\ref{eq:omegatotal}\large]} is comparable with 
experimental results (see~\ref{tab:table2}).

\begin{table}
\caption{\label{tab:table2}Simulated values of Dephasing time at different temperatures.}
\begin{ruledtabular}
\begin{tabular}{lcr}
 & \multicolumn{2}{c}{\em $\tau_{v}$(ps)} \\ \cline{2-3}
Temperature(K) & Simulation & Expt \\\hline 
125.3 & 12.40 & 14.5 \\
140.0 & 19.06 & 23.3 \\
149.3 & 25.13 & 27.8 \\
170.0 & 26.62 & 28.6 \\
186.5 & 26.75 & 26.7 \\ 
\end{tabular}
\end{ruledtabular}
\end{table}

\begin{figure}
\epsfig{file=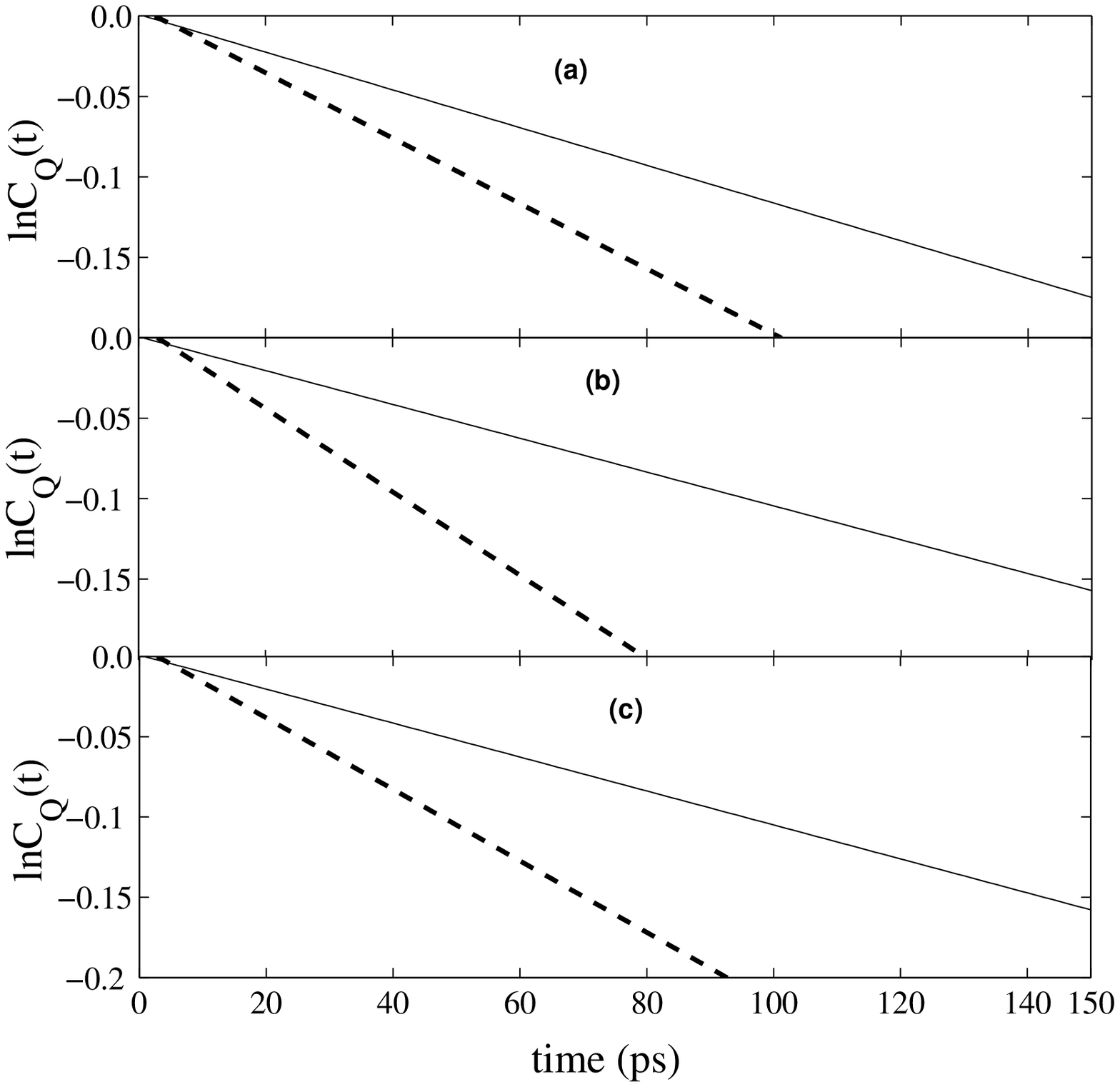,height=12cm,width=14cm}
\caption{\label{fig:FIG5}The normal coordinate time correlation functions 
{\Large[$C_{Q}(t)$\Large]} are plotted against time ($t$) at three temperatures along the 
critical isochore (a) T = 110.8 K, (b) T = 125.3 K, and (c) T = 140.0 K.}
\end{figure}

  In Fig.~$\ref{fig:FIG5}$, a simple line represents the normal coordinate time 
correlation function without including $q$ dependency of LJ parameters whereas the 
dashed line represents the same results with $q$ dependent LJ parameters. 
The former one has a larger correlation time than the latter one. The $C_{Q}(t)$ 
clearly shows the importance of $q$ dependence.  

\begin{figure}
\epsfig{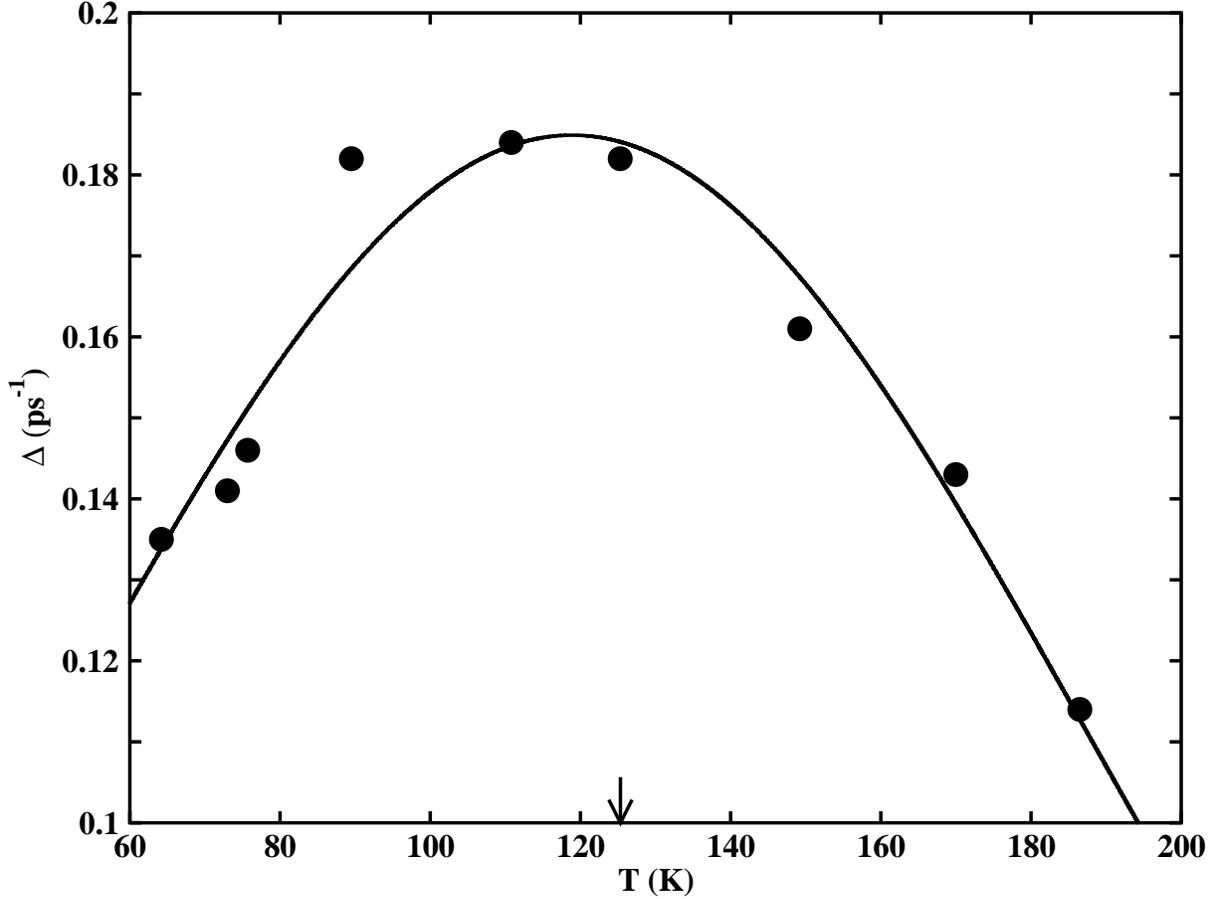}
\caption{\label{fig:FIG6}The rms frequency fluctuation $\Delta$ vs temperature (T) 
along the critical isochore for N$_{2}$. We fitted all 
simulation points with Gaussian. The fitting formula is $y = a[e^{-b(x-c)^
{2}}]$, where $a = 0.184 904, b = 0.000 108 13, and c = 118.874.$}
\end{figure}

  Fig.~$\ref{fig:FIG6}$ displays the root mean square frequency fluctuation $\Delta$ 
{\large[$=\sqrt{\Delta\omega^{2}(0)}$\large]} calculated from unnormalized $C_{\omega}(0)$ at 
different temperatures along the isochore. The calculated $\Delta$ shows a 
nonmonotonic behavior with temperature. $\Delta$ initially increases with 
temperature and shows a maximum near critical temperature indicated by an arrow. 
A sharp decrease is observed for the temperature greater than $T_{c}$. The 
increase in density fluctuation near the critical point is responsible for this 
nonmonotonic behavior of rms frequency fluctuations.

\begin{table}
\caption{\label{tab:table3}Temperature dependence $\lambda$ shaped linewidth 
along the coexistence and critical isochore.}
\begin{ruledtabular}
\begin{tabular}{lcr} 
 Temperature(K) &  Linewidth(Ghz) \\ \hline
     64.2       &     2.764 \\
     71.6       &     2.802 \\
     75.7       &     2.809 \\
     89.5       &     2.812 \\
    110.8       &     2.812 \\
    120.0       &     6.965 \\
    125.3       &    16.12 \\
    135.0       &    12.38 \\
    140.0       &    10.49 \\
    149.2       &     7.96 \\
    170.0       &     7.51 \\
    186.5       &     7.47 \\ 
\end{tabular}
\end{ruledtabular}
\end{table}

\begin{figure}
\epsfig{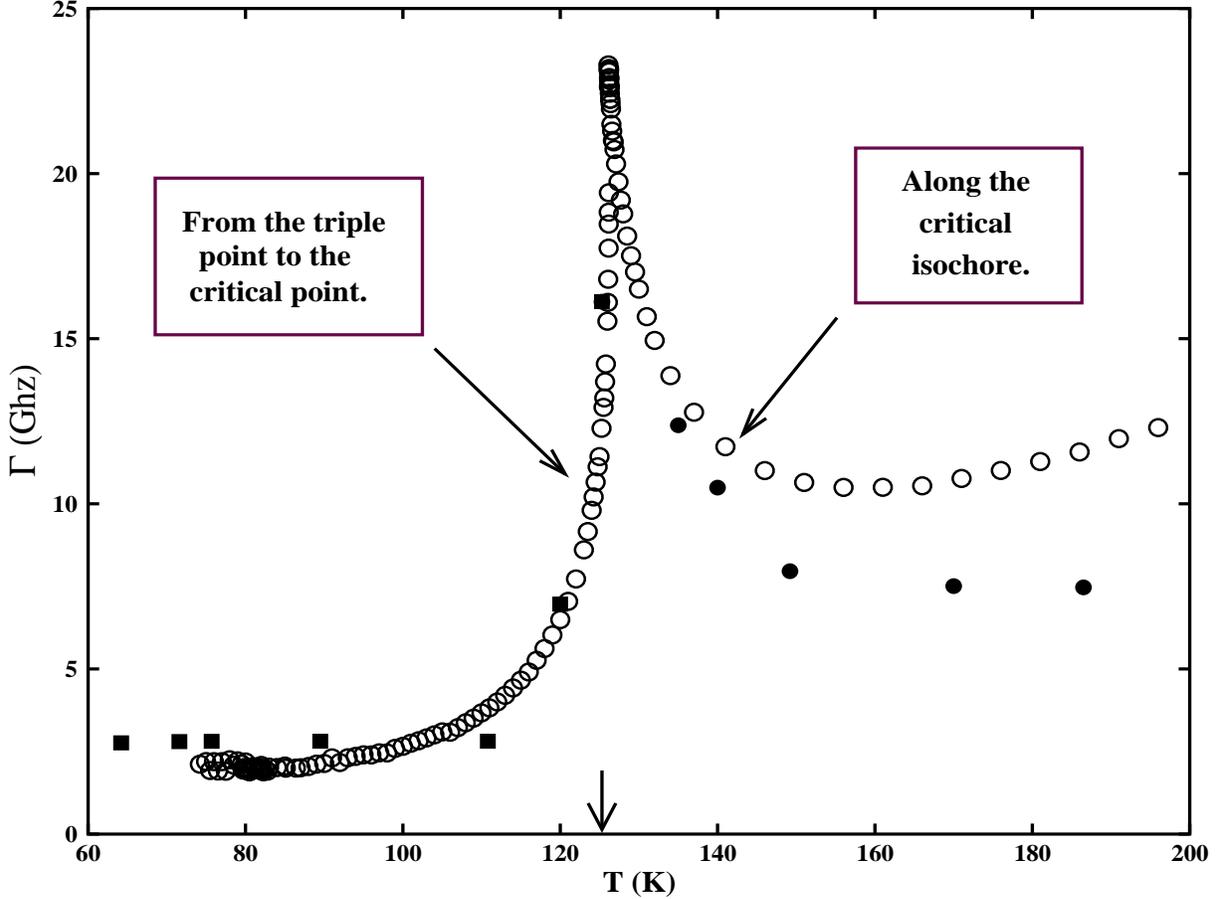}
\caption{\label{fig:FIG7}The lambda ($\lambda$) shaped linewidth ($\Gamma$) 
of nitrogen along the coexistence line (closed square) and the critical isochore 
(closed circle). The experimental results for linewidth along the coexistence line 
 and the critical isochore (open circle) reported by Musso {\it et al}. in 
Ref. \cite{musso} are also shown. The critical point is indicated by an arrow 
on the abscissa.}
\end{figure}

   Linewidth shows an interesting $\lambda$-shaped feature when it is plotted 
against temperature (see~\ref{tab:table3}) for the two different branches of 
calculation for N$_{2}$ as shown in Fig.~$\ref{fig:FIG7}$. This figure is very similar to 
the one observed in experiment (see Fig.~$\ref{fig:FIG4}$ of Ref \cite{musso}). It is 
interesting to note the sharp rise in the dephasing rate as the critical 
point is approached. There are six main  contributions, (a) density 
{\large[$C^{\rho}_{\omega}(t)$\large]}, (b) VR coupling {\large[$C^{VR}_{\omega}(t)$
\large]}, (c) resonance {\large[$C^{Rs}_{\omega}(t)$\large]}, (d) density-VR coupling 
{\large[$C^{\rho-VR}_{\omega}(t)$\large]}, (e) density-resonance {\large[$C^{VR-Rs}_{
\omega}(t)$\large]}, and (f) VR-resonance {\large[$C^{\rho-Rs}_{\omega}(t)$\large]} 
which are responsible for the sharp rise in total linewidth near 
the critical point. These are the time integrals of diagonal and cross-terms 
in the frequency fluctuation time correlation function. We shall come back to 
this point a bit later.

\begin{figure}
\epsfig{file=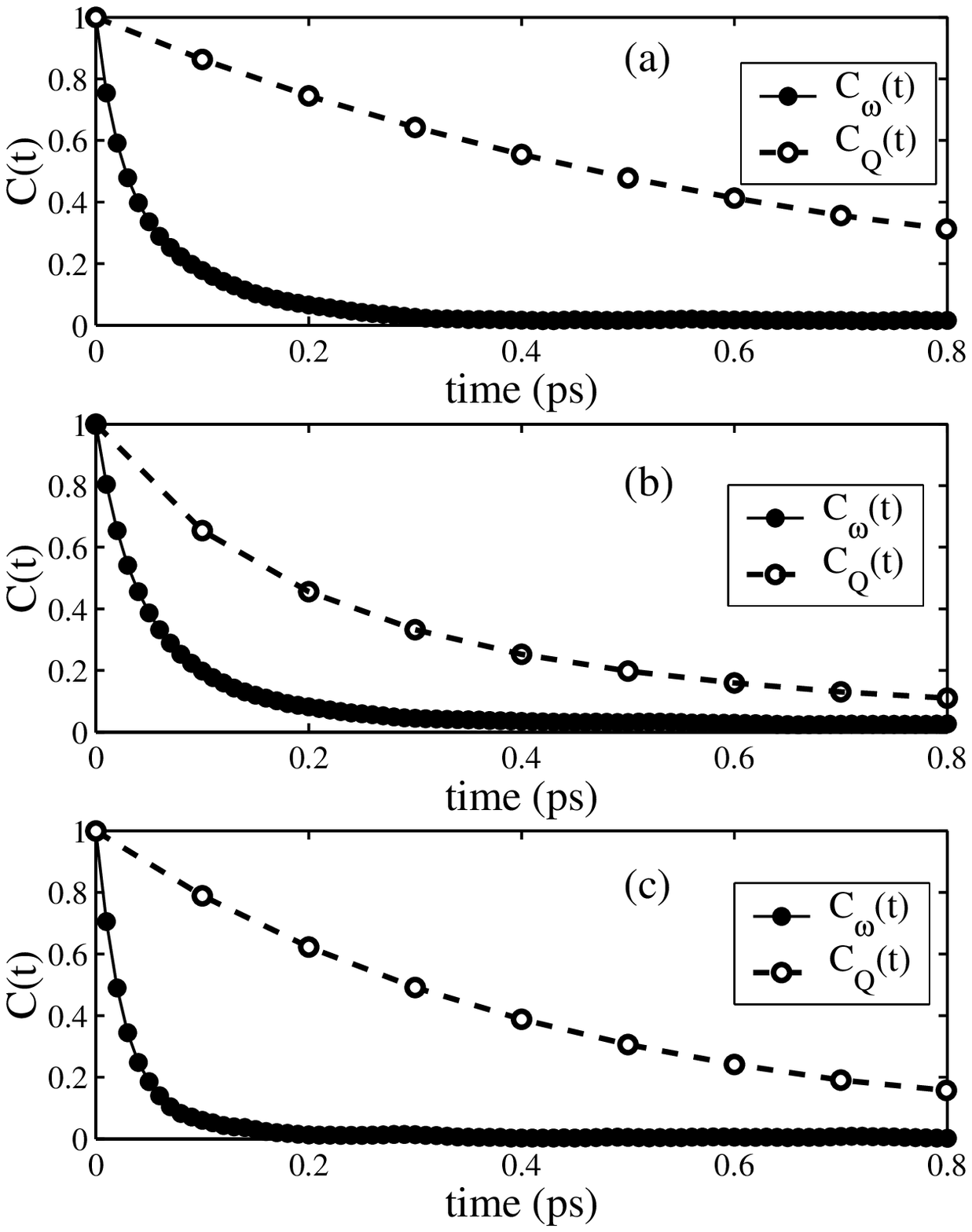}
\caption{\label{fig:FIG8}The frequency fluctuation time correlation function, 
$C_\omega(t)$, and the normal coordinate time correlation function, $C_Q(t)$, are 
plotted at (a) 186.5 K, and (b) 140.0 K (along critical isochore), and (c) the 
$C_\omega(t)$ and $C_Q(t)$ are plotted along the coexistence line at temperature 
75.7 K.}
\end{figure}

  A crossover from Lorentzian-type line shape to Gaussian line shape is found 
to take place when there is a large separation in the time scales of decay 
of $C_{\omega}(t)$ and $C_{Q}(t)$ \cite{musso,oxrev2} cease to exist and the 
two time correlation functions begin to overlap. We have calculated $C_{\omega}
(t)$ and $C_{Q}(t)$ at three temperatures near the critical point (see 
Fig.~$\ref{fig:FIG8}$). The decay of $C_{\omega}(t)$ becomes significantly 
faster, reducing the gap of decay between the two correlation functions. Indeed, 
the computed line shape becomes Gaussian near the critical point but otherwise 
remains Lorentzian-type both above and below the critical temperature. Note 
that the frequency modulation time correlation function decays fully in about 
200 fs.

  To understand the origin of this critical behavior, we have carefully analyzed 
each one of the six terms which consist of three autocorrelations and three 
cross-terms between density, vibration-rotation coupling, and resonance terms 
which have been mentioned earlier. 

\begin{figure}
\epsfig{file=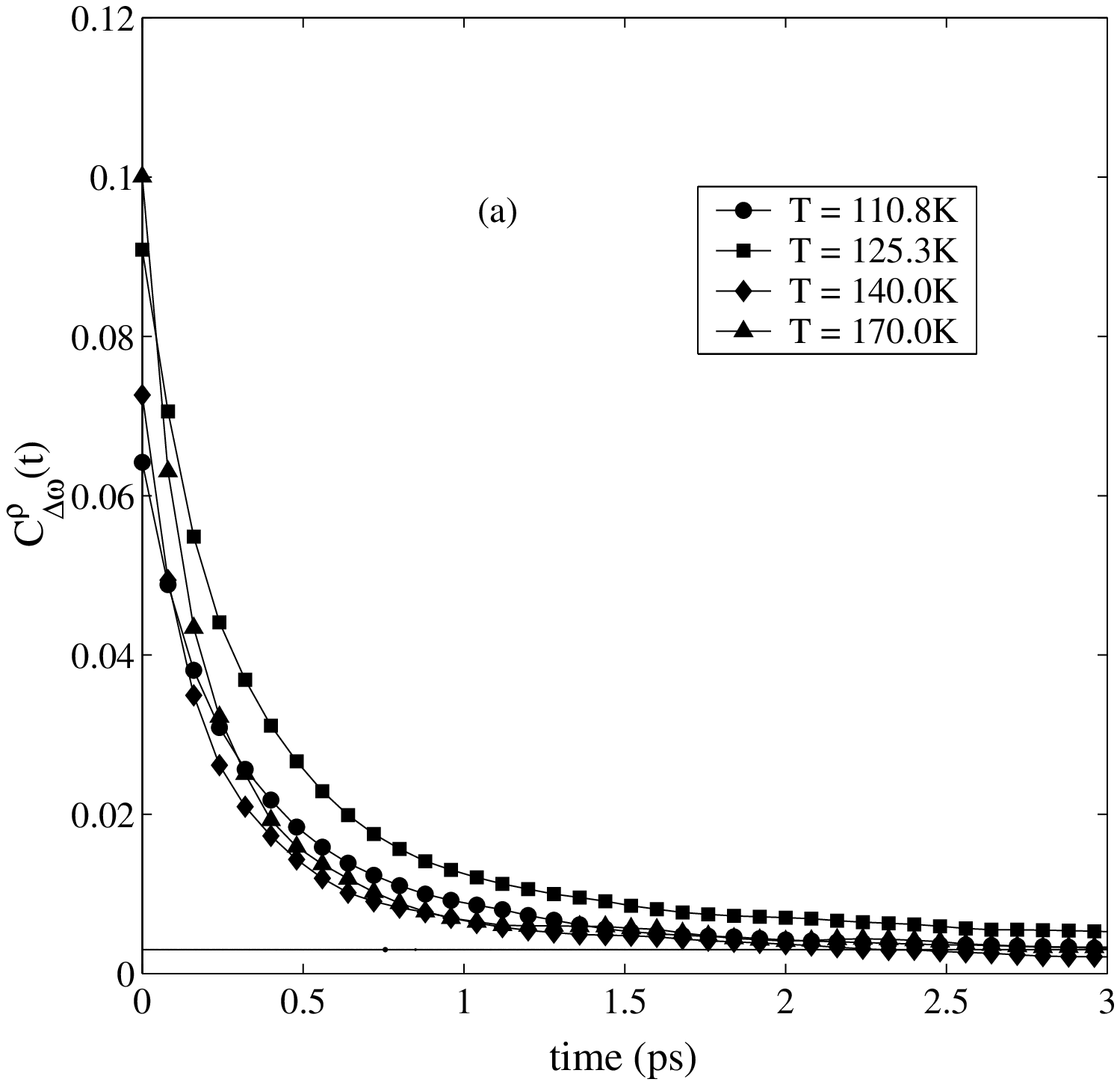,height=10cm,width=12cm}\\
\vspace*{2cm}
\epsfig{file=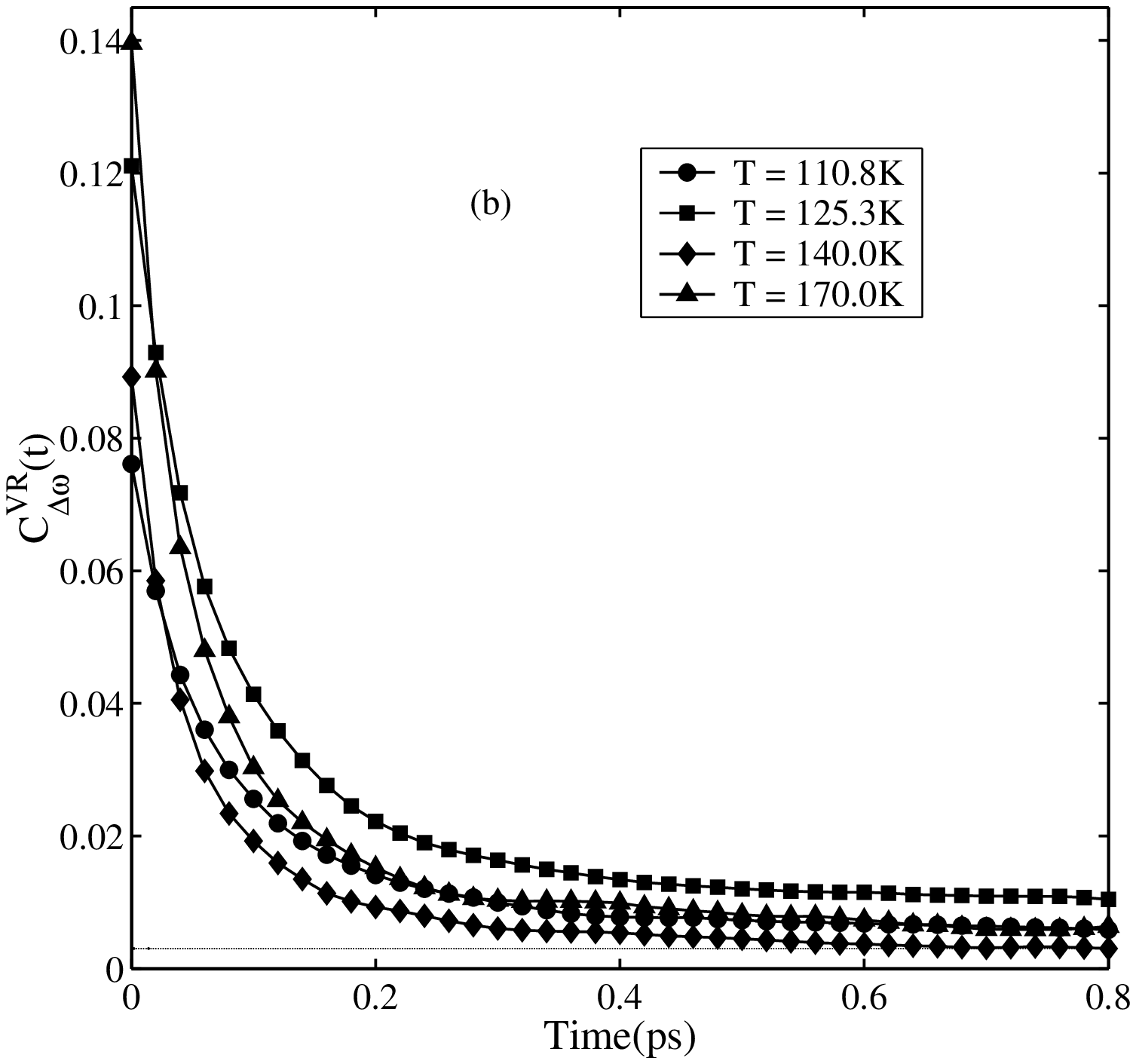,height=10cm,width=12cm}
\end{figure}
\begin{figure}
\epsfig{file=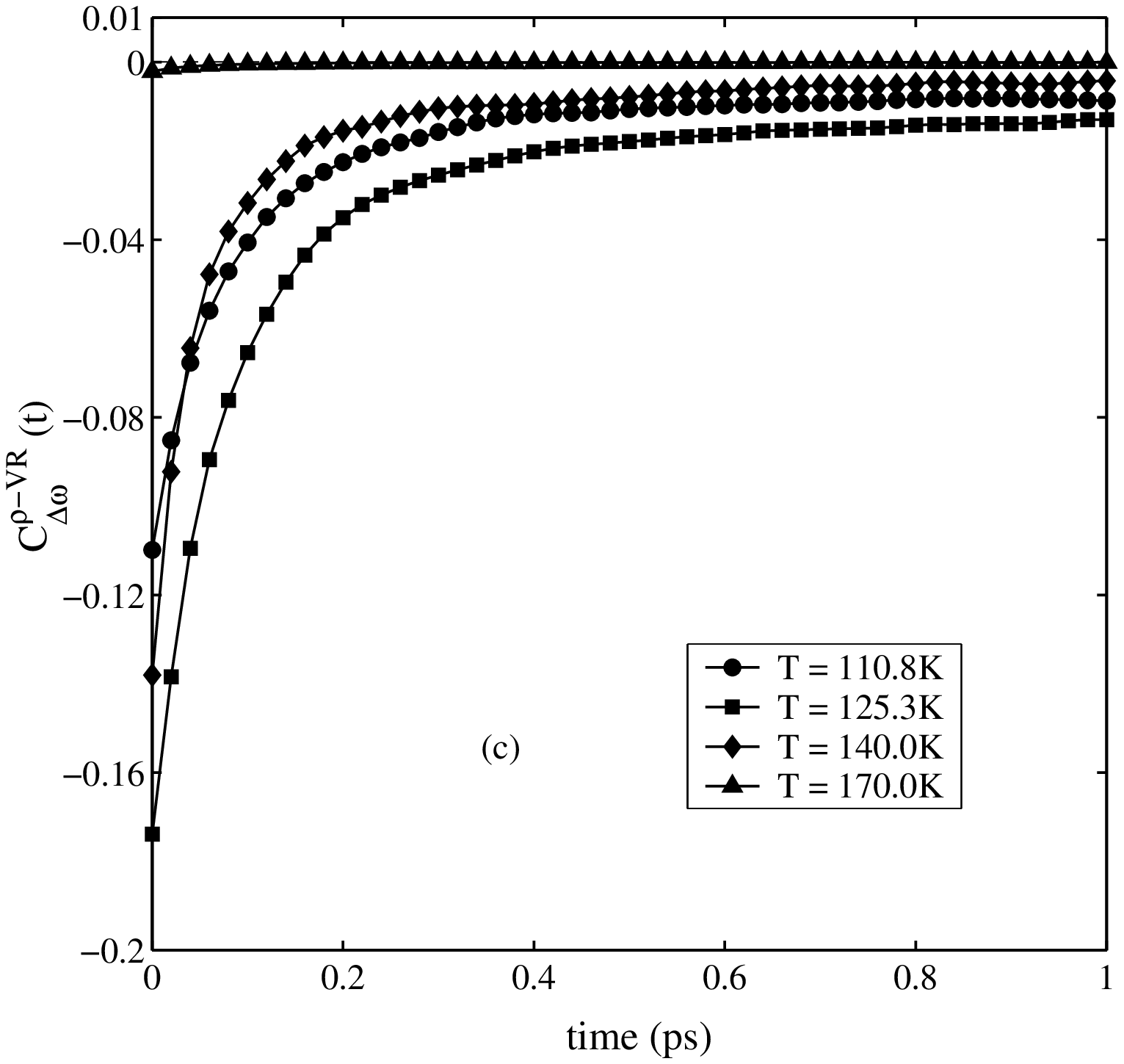,height=8cm,width=10cm}\\
\vspace*{2cm}
\epsfig{file=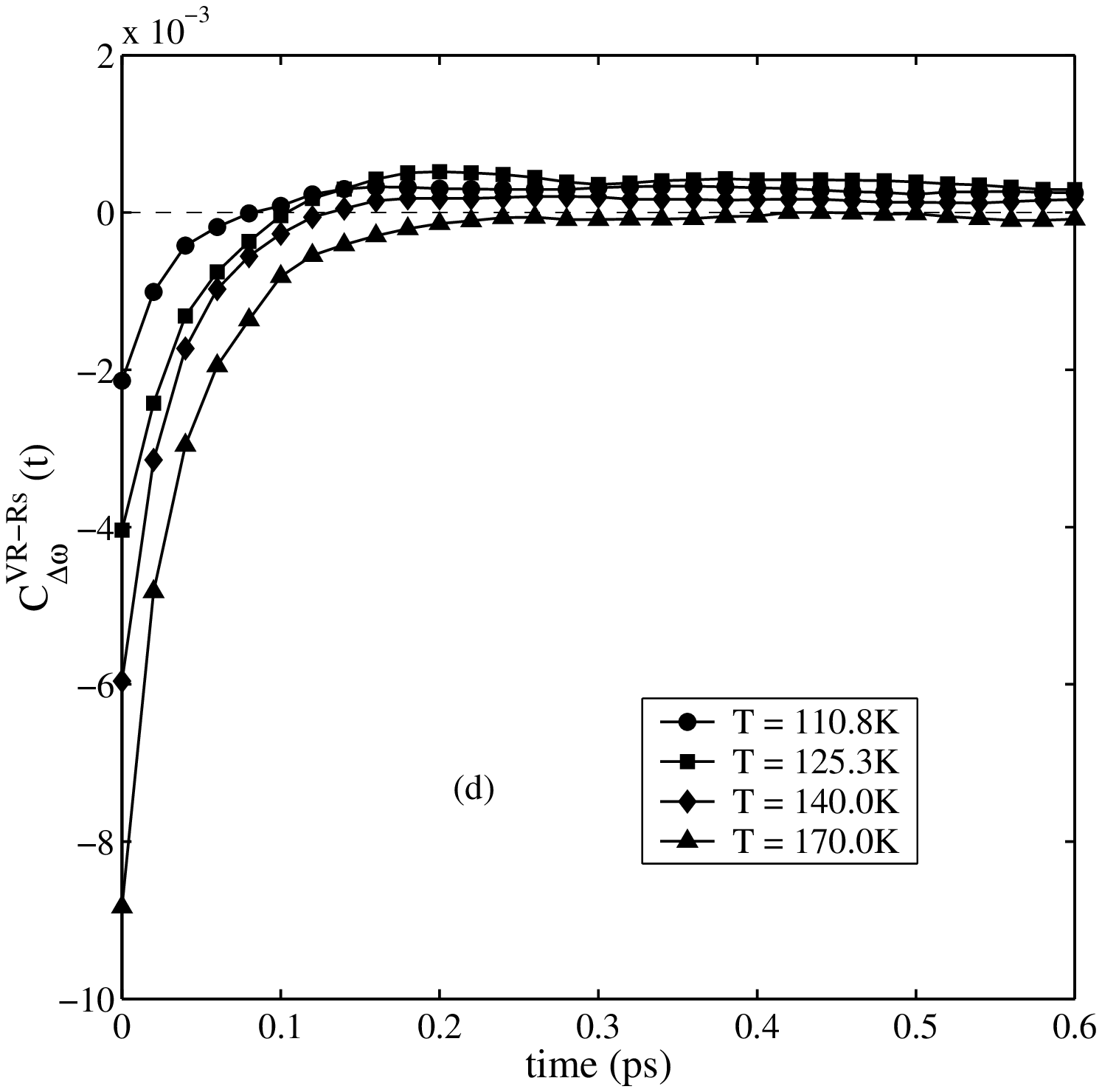,height=8cm,width=10cm}
\caption{\label{fig:FIG9}The decopositions of unnormalized frequency 
fluctuation time correlation functions for N$_{2}$ into (a) density-density, 
(b) VR-VR coupling, (c) density-VR and (d) VR-resonance at four temperatures 
along the critical isochore. The density-resonance and resonance-resonance terms 
are relatively small and are not shown here.}
\end{figure}

  Fig.~$\ref{fig:FIG9}$ shows the time dependence of the four dominating terms , the two 
autocorrelations {\large[Fig.~$\ref{fig:FIG9}$(a) and Fig.~$\ref{fig:FIG9}$(b)\large]} and two 
cross-correlations (Fig.~$\ref{fig:FIG9}$(c) and Fig.~$\ref{fig:FIG9}$(d)). 
The decompositions of line shift, linewidth, and the temperature dependent 
quantities are fully dependent on the contributions which come from all six terms. 
As we approach the critical point along the critical isochore, the magnitude 
of contributions from different terms are increased. Results for only four 
different temperatures including the critical point are shown in 
Fig.~$\ref{fig:FIG9}$. All the remaining terms, resonance-resonance and density-resonance, 
are found to be unimportant in comparison with these four terms.

\begin{figure}
\epsfig{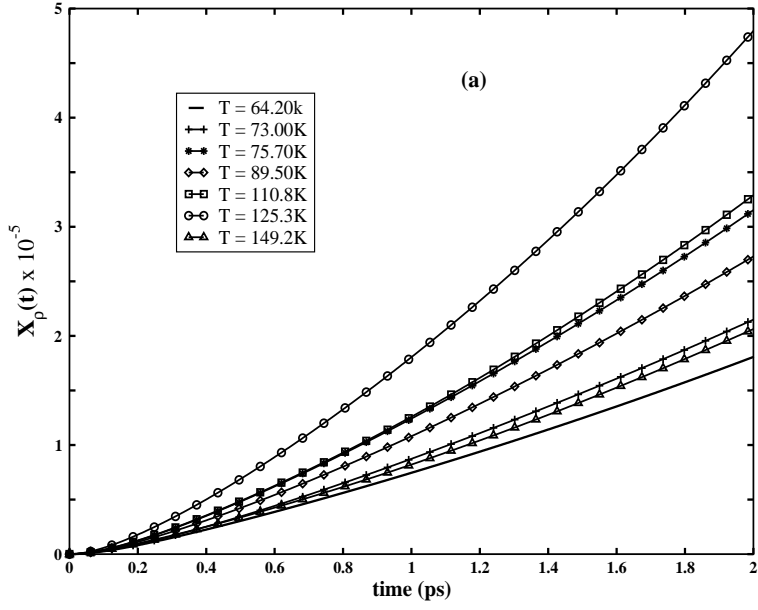}\\
\vspace*{2cm}
\epsfig{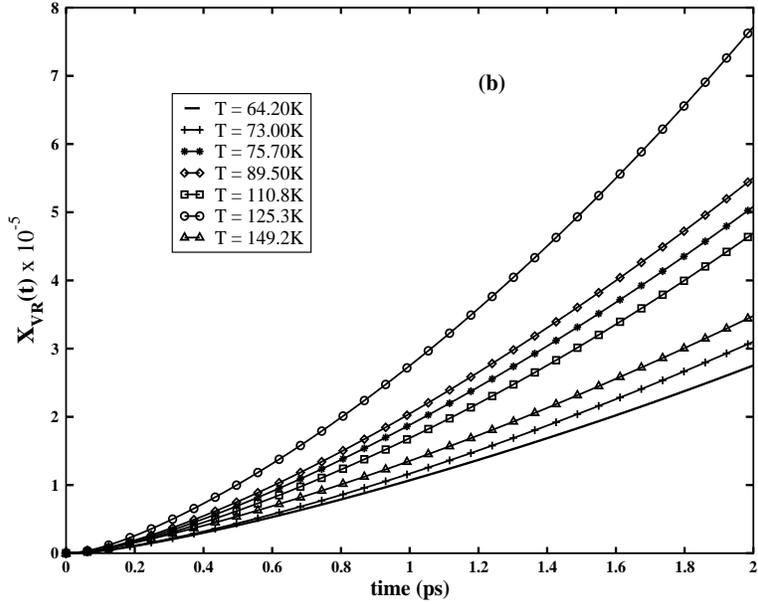}
\caption{\label{fig:FIG10}The time dependence of the relative contributions 
of the (a) density-density, $X_{den}(t)$, and (b) VR coupling, $X_{VR}(t)$, 
terms at different temperatures along the critical isochore.}
\end{figure}

  The time dependence relative contribution $X_{ij}(t) = \int_{0}^{t}dt^{\prime}
(t-t^{\prime})C_{\omega}^{ij}(t^{\prime})$ of the density {\large[Fig.~$\ref{fig:FIG10}$a\large]} 
and vibration-rotation coupling term (Fig.~$\ref{fig:FIG10}$b) are plotted for 
seven state points along the critical isochore, which are found to be dominant 
near the critical point. The sharp {\it rise} in the value of the integrand as the 
critical temperature is approached and the {\it fall} when it is crossed. We have 
found that both these contributions at the critical point are distinct compared 
to the other state points. Thus the rise and fall of the dephasing rate arise 
partly from the rise and fall in the density and the vibration-rotation terms. 
We have calculated the slopes of the relative contributions by the linear 
fitting in the long time at a particular temperature $T$ = 125.3 K. The values of 
the slope are $5.4\times 10^{-5}$, $9.5\times10^{-5}$ and $-6.8\times10^{-5}$ 
for density-density, VR-VR coupling, and density-VR terms, respectively. The major 
contribution comes from the VR coupling term among the four dominating terms.

\section{Dynamical heterogeneities near the critical point.}

\begin{figure}
\epsfig{file=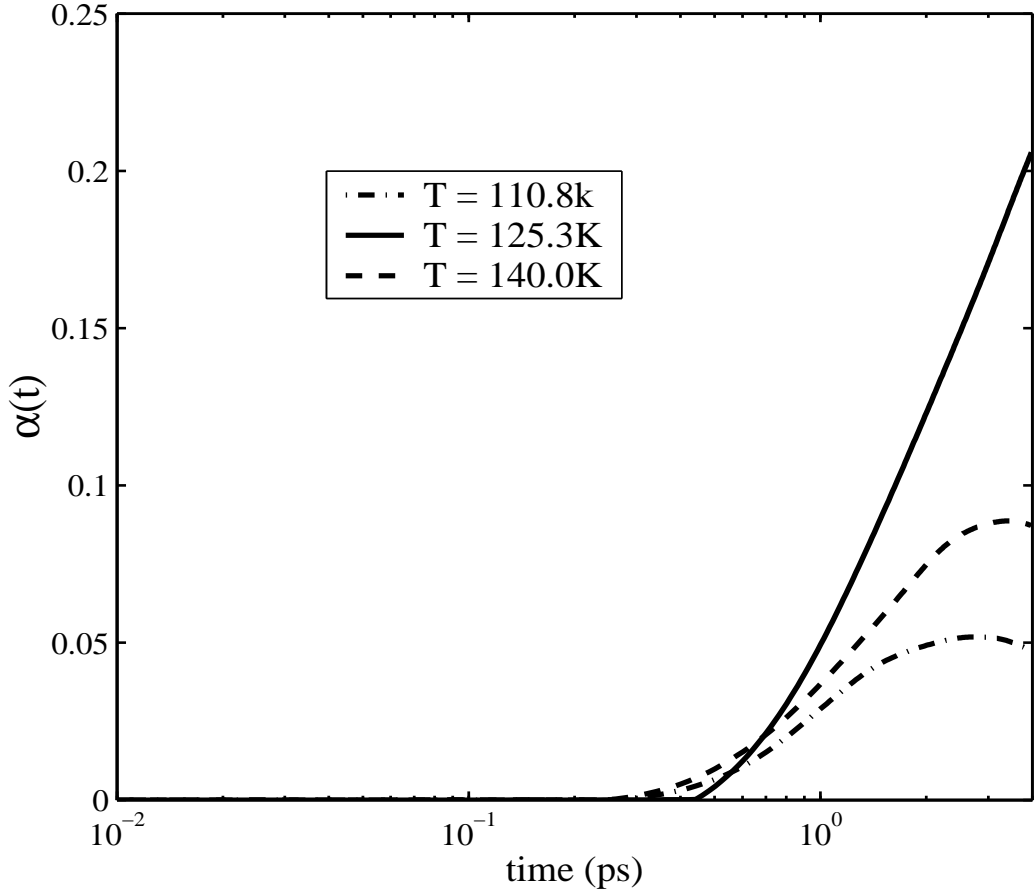,height=12cm,width=14cm}
\caption{\label{fig:FIG11}The non Gaussian parmeter $\alpha(t)$ is plotted 
against time ($t$) at three different temperatures along the critical isochore.}
\end{figure}

  We have investigated the presence of dynamical heterogeneities in the fluid 
to further explore the origin of these anomalous critical temperature effects
 \cite{stanley,ma}. This can be quantified by the well-known non-Gaussian 
parameter $\alpha(t)$. It is defined as \cite{ara}
\begin{equation}
\alpha(t) = \left(\frac{3}{5}\right)\frac{<r^{4}(t)>}{<r^{2}(t)>^{2}} - 1,
\end{equation}
where $\large<\Delta r(t)^{2}\large>$ is the mean squared displacement and $\large<\Delta 
r(t)^{4}\large>$ the mean quartic displacement of the center of mass of the nitrogen 
molecule. It can only approach zero (and hence Gaussian behavior) at a time
scale larger than the time scale required for individual particles to sample 
their complete kinetic environments. The function $\alpha$(t) is large near 
the critical point at times 0.5-5 ps as observed in Fig.~$\ref{fig:FIG11}$ which 
indicates the presence of long lived heterogeneities near $T_{c}$.

\section{Negative cross-correlation between the density and VR coupling term}

  The role of cross-terms (which can be negative) are extremely important for
the decay of $C_{\omega}(t)$ which occurs in the femtosecond time scale along the
critical isochore. We have calculated the cross-terms among density, VR coupling, 
and resonance term. The cross-terms of VR coupling with density and resonance are 
found to be negative {\large[see Figs.~$\ref{fig:FIG9}$(c) and $\ref{fig:FIG9}$(d)\large]}. 
As mentioned earlier the major contributions to the frequency fluctuation come from 
the density as well as from the VR coupling term. One of the main reasons for 
ultrafast decay of $C_{\omega}$ is the cancellations of negative cross-terms from 
the total contribution.

\begin{figure}
\epsfig{file=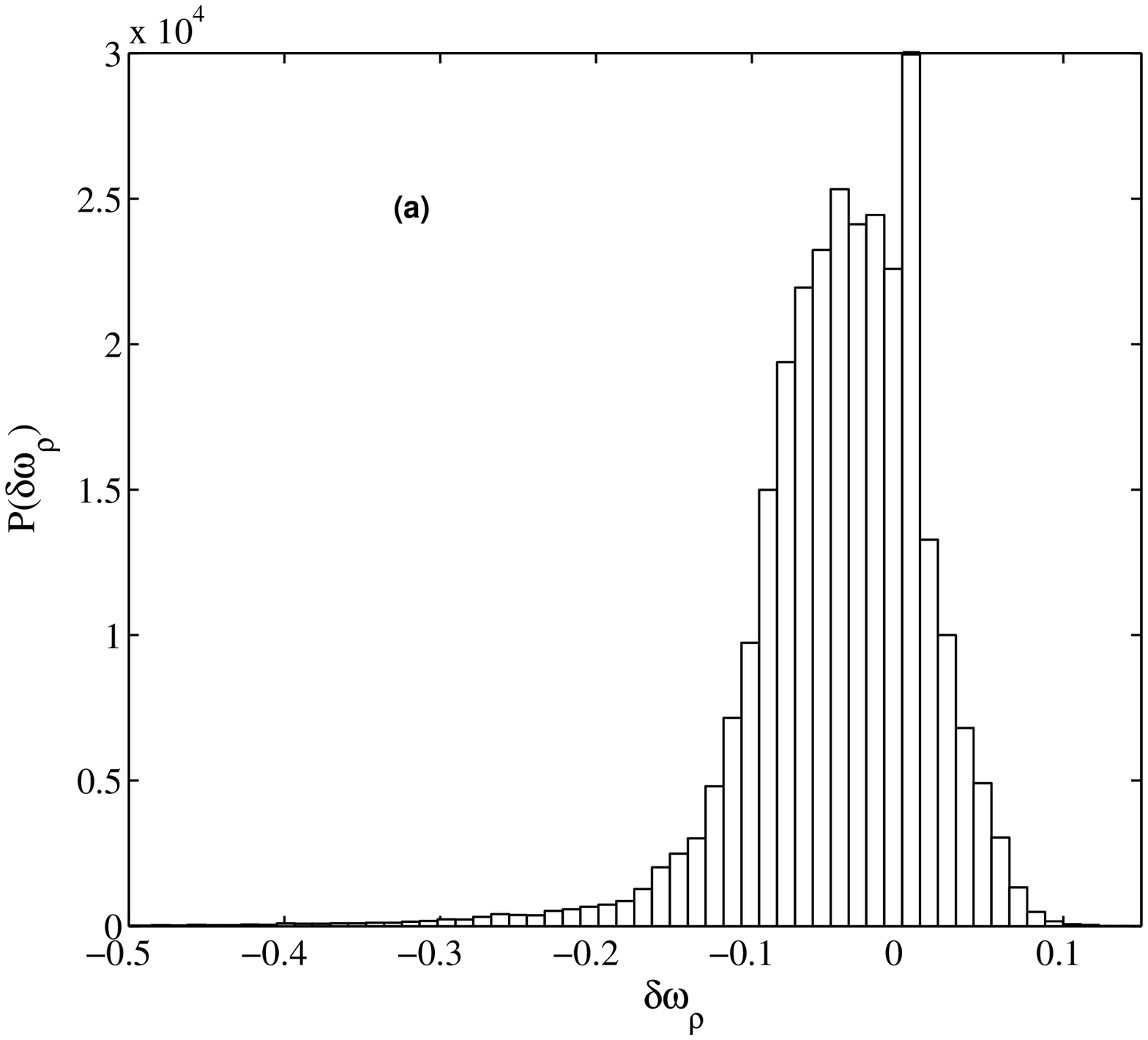,height=10cm,width=12cm}
\epsfig{file=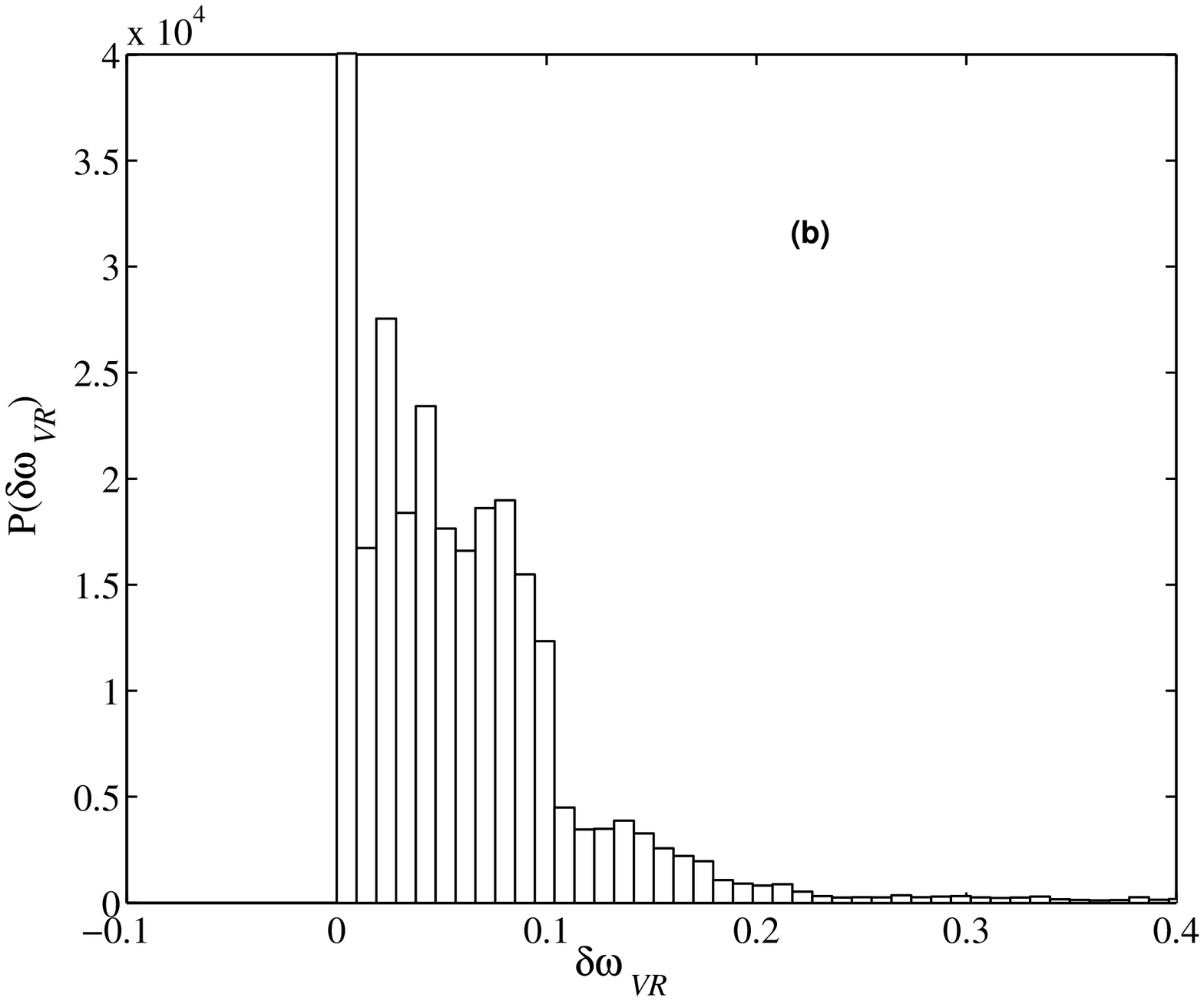,height=10cm,width=12cm}
\caption{\label{fig:FIG12}The distribution of fluctuation in (a) the density 
term, and (b) the VR coupling contribution to total fluctuation in the frequency at the 
temperature 110.8 K.}
\end{figure}

\begin{figure}
\epsfig{file=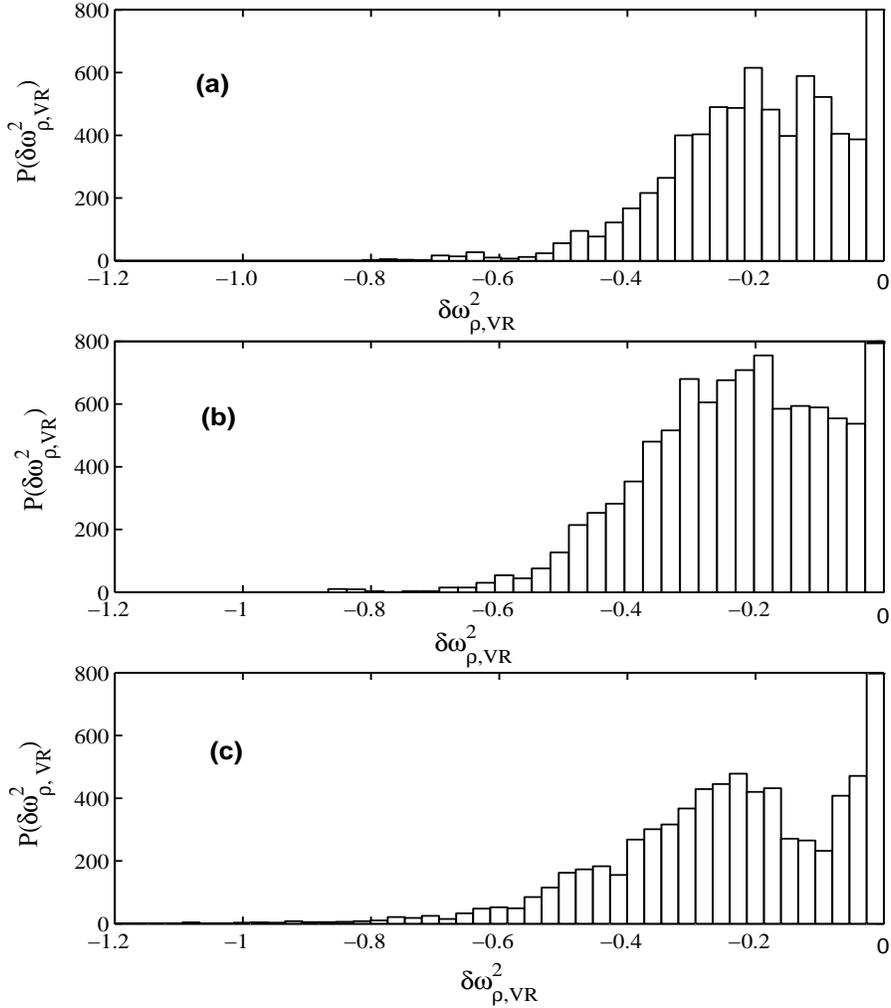,height=14cm,width=12cm}
\caption{\label{fig:FIG13}The distribution of correlation between density-VR 
coupling at three temperatures, (a) 110.8 K, (b) 125.3 K and (c) 140.0 K, 
respectively. Note that this figure is entirely different from Fig.~$\ref{fig:FIG14}$.}
\end{figure}

   Now the question is why are these cross-terms are negative? The distribution of 
density dependent and the VR coupling dependent part of the total frequency 
fluctuation along the isochore are shown in Figs.~$\ref{fig:FIG12}$(a) and $\ref{fig:FIG12}$(b), 
respectively. The peak of those distributions is near zero $\delta\omega_{VR}$ which proves 
that the homogeneity condition is satisfied along the isochore. It is evident that the 
distribution of $\delta\omega_{VR}$ is always positive while the distribution of 
$\delta\omega_{\rho}$ showed the long negative tail. The distribution of the 
product of density and VR coupling terms are plotted for different temperatures 
in Fig.~$\ref{fig:FIG13}$. It is clearly seen that this distribution is negative. A detailed 
analysis of the origin of the negative sign of the cross-correlation between 
AA and VR coupling terms is given in Appendix B. On the other hand 
$\Delta\omega_{\rho}$ is directly proportional to the force acting on the bond 
of the diatoms. If force $F$ is large, the velocity will be large and it reflects 
that the $J^{2}$ will be decreased when $F$ is increased and vice versa. This is 
the origin of anticorrelation between the density and VR coupling terms in 
$C_{\omega}(t)$.
 
\begin{figure}
\epsfig{file=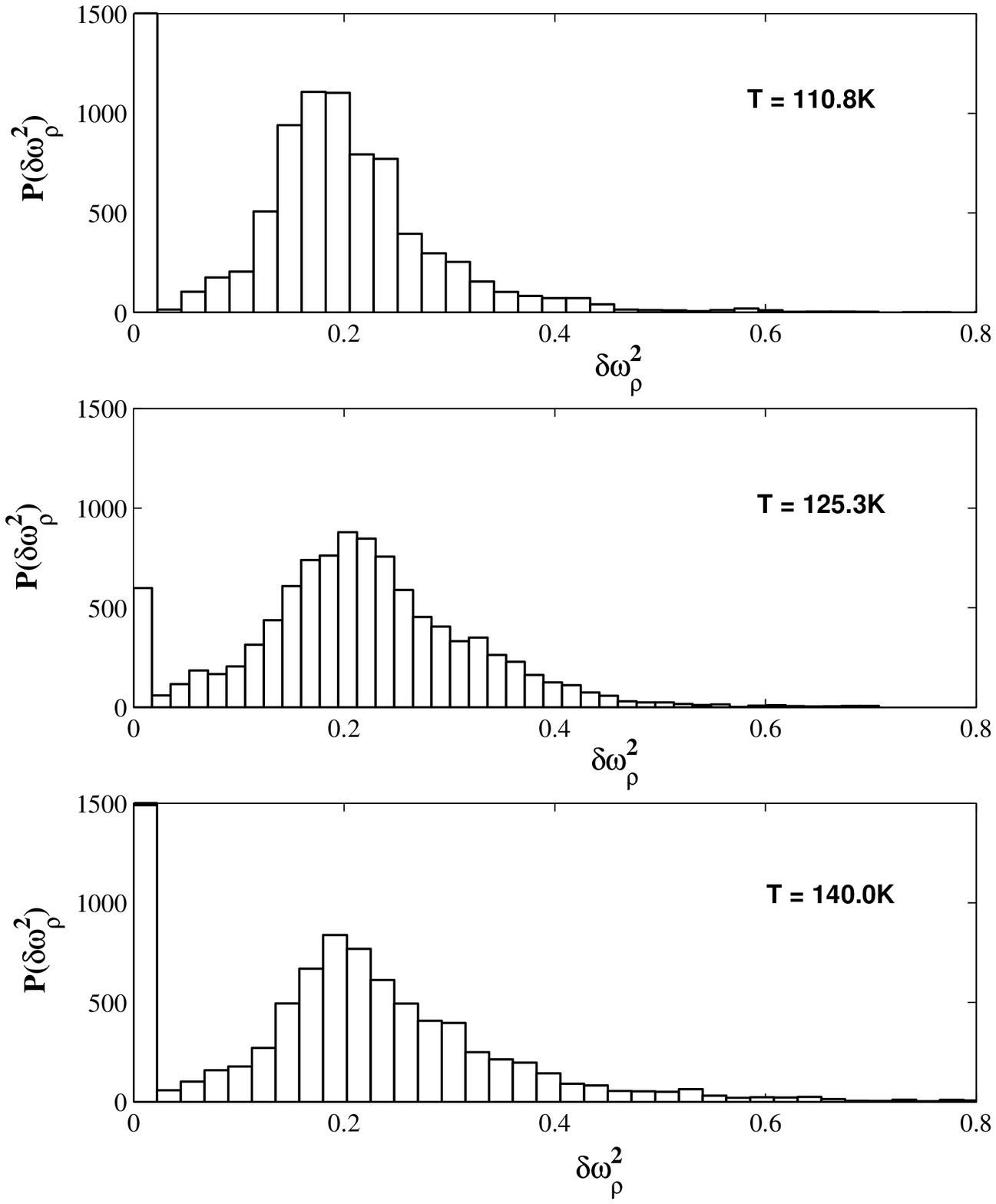,height=10cm,width=8cm}
\epsfig{file=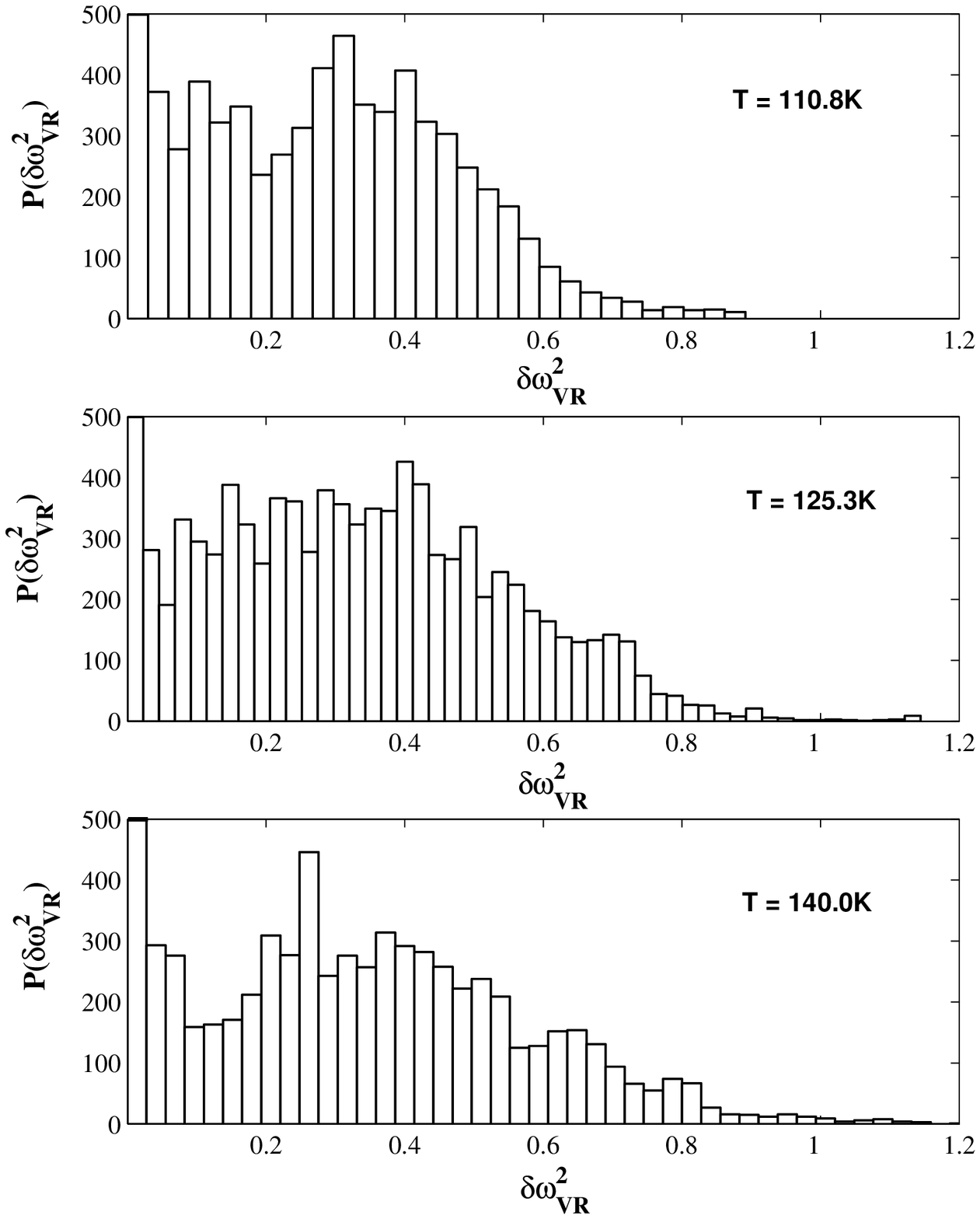,height=10cm,width=8cm}
\caption{\label{fig:FIG14}The distribution of (a) the density-density term 
and (b) the VR-VR coupling at term three temperatures 110.8 K, 125.3 K and 140.0 K 
respectively.}
\end{figure}

  The distribution of correlation between the density-density term and the VR-VR coupling 
term in the frequency fluctuation time correlation function are shown in 
Figs.~$\ref{fig:FIG14}$(a) and $\ref{fig:FIG14}$(b), respectively. Both distributions are 
positive and the homogeneity condition of frequency fluctuations is satisfied in both 
cases.

\section{Mode Coupling Theory Analysis.}

  We can use mode coupling theory to obtain an expression for the atom-atom 
contribution to the frequency modulation time correlation function. The main 
steps have already been discussed by Gayathri {\it et al}. \cite{gay3} and need not be 
repeated here. In brief, MCT gives the following expression for the density 
dependent frequency modulation time correlation function \cite{sarika}
considering that the number density is the only relevant slow variable in dephasing, 
\begin{widetext}
\begin{eqnarray}
\left<\Delta\omega_{\rho}(0)\Delta\omega_{\rho}(t)\right>=\frac{(k_{B}T)^{2}}
{6\pi^{2}\hbar^{2}\rho}\int_{0}^{\infty}k^{2}dkF_{s}(k,t)c^{2}(k)F(k,t).
\end{eqnarray}
\end{widetext}
Where $c(k)$ is the Fourier transform of the two particle direct correlation 
function. The main contribution is derived from the long wavelength 
(that is small $k$) region near the critical point (CP). In the small $k$ limit the 
hydrodynamic expressions for two-point correlation functions are given by 
\begin{subequations}
\begin{eqnarray}
F_{s}(k, t)= e^{-D_{s}k^{2}t},\label{stfra}
\end{eqnarray}
\begin{eqnarray} 
F(k,t)= S(k)e^{-D_{T}k^{2}t},\label{stfrb}
\end{eqnarray}
\end{subequations}
where $F_{s}(k,t)$ is the self-intermediate scattering function and $F(k,t)$
is the intermediate scattering function. Here $D_{s}$ is the self-diffusion 
coefficient, $S(k)$ is the static structure factor, and $D_{T}$ is the thermal 
diffusivity. Thus $\left<\Delta\omega_{\rho}(0)\Delta\omega_{\rho}(t)\right> 
\simeq S(k \rightarrow 0)e^{-(D_{s}+D_{T})k^{2}t}$. Near the CP, $S(k\rightarrow 0)$ 
becomes very large ( as compressibility diverges at $T = T_{c}$ ), leading 
towards a Gaussian behavior for line shape. $D_{T}$ also undergoes a slowdown 
near $T_{c}$. However, a limitation of the above analysis is the absence of the 
VR term which contributes significantly and may mask some of the 
critical effects.\\ 

\begin{figure}
\epsfig{file=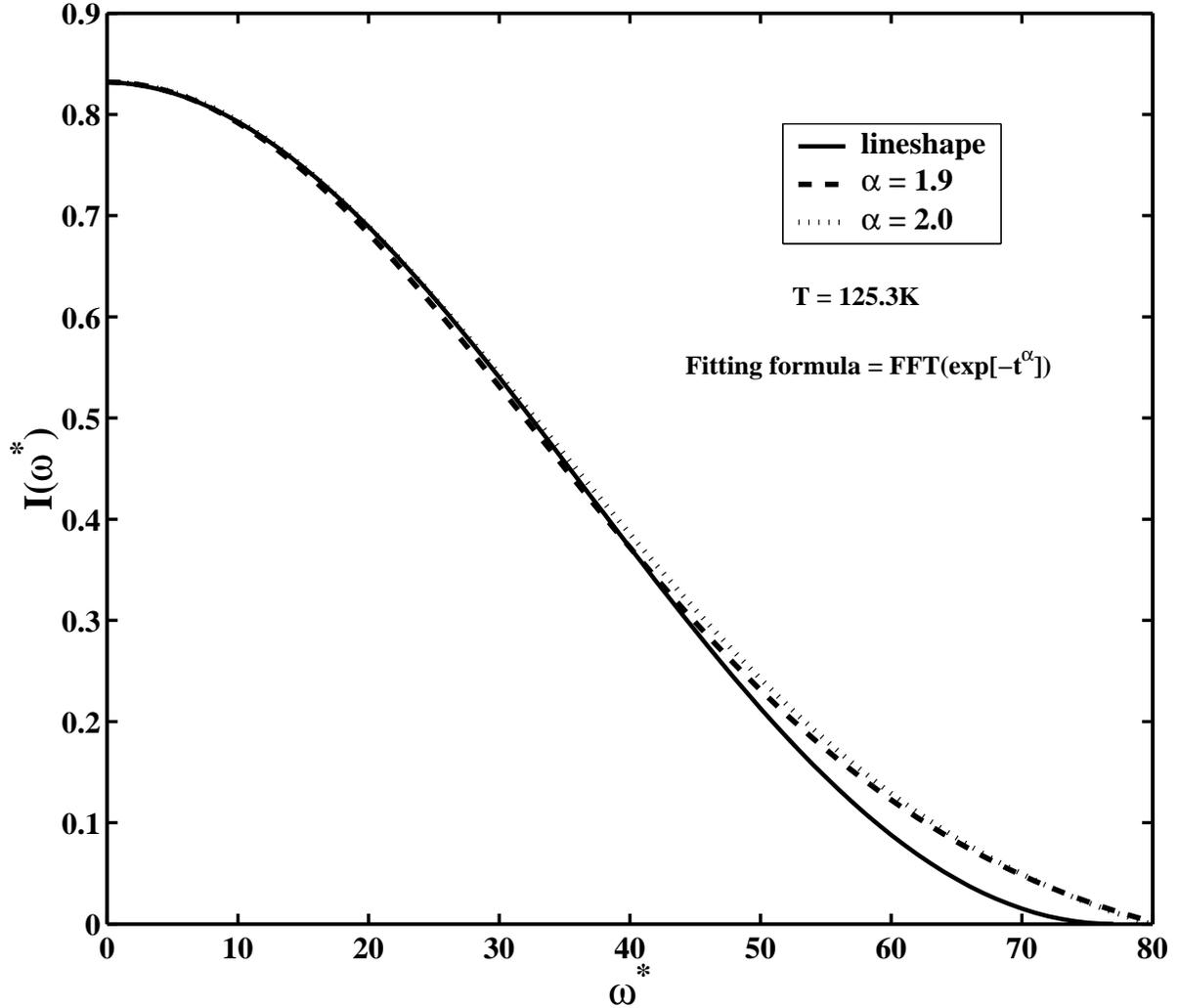,height=14cm,width=16cm}
\caption{\label{fig:FIG15}The behavior of line shape at the temperature 123.5 K.}
\end{figure}

    All these lead to a large value of $\large<\Delta\omega^{2}(0)\large>$, which may 
lead to a Levy distribution from time dependence of $\left<Q(0)Q(t)\right>$ as 
discussed earlier by Mukamel {\it et al}. \cite{muk}. At high temperature, the latter 
dominates over the density term. We have calculated the line shape and fitted with 
the Fourier transform of Levy distribution ( see Fig.~$\ref{fig:FIG15}$). 
Clearly as $T_{c}$ is approached, the low $k$ fluctuations become more important
and, if the high $k$ contributions are sufficiently small, they will dominate 
the line shape. We would like to mention that the line shape becomes Gaussian
(see the Fig.~$\ref{fig:FIG15}$) instead of the Fourier transform of Levy 
distribution near the critical point. The small $k$ picture described here is quite 
from the usual collisional broadening picture of dephasing. A complete Lorentzian 
behavior is predicted only in the low temperature liquid phase. Interestingly, 
the predicted divergence of $\large<\Delta\omega^{2}(0)\large>$ very close to $T_{c}$ 
enhances the rate of dephasing and this shifts the decay of $\large<Q(t)Q(0)\large>$ to 
short times, which gives rise to the Gaussian behavior.

  We next use MCT to analyze the cross-correlation between VR coupling and the 
atom-atom ($\Delta\omega_{\rho}$) term. According to density functional theory 
(DFT) the effective potential energy ($V_{eff}$) of the mean field force $\left[{\bf F}
= -{\bf\nabla}V_{eff}({\bf r},t)\right]$ can be written in terms of the two-particle 
correlation function $c({\bf r})$ without including orientation(${\bf\Omega}$) 
as
\begin{eqnarray}
\beta V_{eff}({\bf r},t) = - \int d{\bf r^{\prime}}c({\bf r}-{\bf r^{\prime}})
\delta\rho({\bf r^{\prime}},t).
\end{eqnarray}
Assuming that the density term is approximated by the isotropic limit, we can 
therefore write the atom-atom terms as
\begin{eqnarray}
\Delta\omega_{\rho}({\bf r},t) = -k_{B}T/\hbar\int d{\bf r^{\prime}}c({\bf r}-
{\bf r^{\prime}},t)\delta\rho({\bf r^{\prime}},t).
\end{eqnarray}
The calculation of the VR term is a bit more complicated. The angular momentum 
${\bf J}(t)$ at time $t$ can be expressed in terms of torque ${\bf N}(t^{\prime})$
at earlier times ($t^{\prime}>t$), 

\begin{widetext}
\begin{subequations}
\begin{eqnarray}
{\bf J}(t) = {\bf J}(0)+ \int dt^{\prime}{\bf N}(t^{\prime}).
\end{eqnarray}
 From Eq.~\ref{eq:omegavr2} of the Appendix B, the VR coupling can be written as
\begin{eqnarray}
\Delta\omega_{VR}({\bf r},{\bf\Omega},t)=A{\bf J}({\bf r},{\bf\Omega},t)\cdot
{\bf J}({\bf r},{\bf\Omega},t)
=A\large[J^{2}({\bf r},{\bf\Omega},0)+2\int dt^{\prime}{\bf J}({\bf r},
{\bf\Omega},0)\cdot{\bf N}({\bf r},{\bf\Omega},t^{\prime})\nonumber\\+\int dt^
{\prime}\int dt^{\prime\prime}{\bf N}({\bf r},{\bf\Omega},t^{\prime})\cdot
{\bf N}({\bf r},{\bf\Omega},t^{\prime\prime})\large].
\end{eqnarray}
\end{subequations}
\end{widetext}
Here $A$ is a constant (see Appendix B) and $\delta\rho({\bf r^{\prime}},
{\bf\Omega^{\prime}},t^{\prime})$ is the fluctuation in the number density.

 The expression for the cross-correlation can be expressed as

\begin{widetext}
\begin{eqnarray}
\left<\Delta\omega_{VR}({\bf r},{\bf\Omega},t)\Delta\omega_{\rho}({\bf r},0)\right>=-
A(k_{B}T/\hbar)\large[\int d{\bf r^{\prime}}c({\bf r}-{\bf r^{\prime}})\left<J^{2}
({\bf r},{\bf\Omega},0)\delta\rho({\bf r^{\prime}},0)\right>\nonumber \\+ 2\int 
dt^{\prime}\int d{\bf r^{\prime}}c({\bf r}-{\bf r^{\prime}})\left<{\bf J}({\bf r},
{\bf\Omega},0)\cdot{\bf N}({\bf\Omega},t^{\prime})\delta\rho({\bf r^{\prime}}
,0)\right>\nonumber\\ +\int dt^{\prime}\int dt^{\prime\prime}\int d{\bf r^{\prime}}
c({\bf r}-{\bf r^{\prime}})\left<{\bf N}({\bf\Omega},t^{\prime})\cdot{\bf N}({\bf
\Omega},t^{\prime\prime})\delta\rho({\bf r^{\prime}},0)\right>\large].
\label{eq:crossterm}
\end{eqnarray}
By using DFT again, the torque (${\bf N}$) can expressed in terms of density
functions as
\begin{eqnarray}
{\bf N}({\bf r},{\bf\Omega},t) = {\bf\nabla}_{\bf\Omega}\int d{\bf r^{\prime}}
d{\bf\Omega^{\prime}}c({\bf r}-{\bf r^{\prime}},{\bf\Omega},{\bf\Omega^{\prime}})
\delta\rho({\bf r^{\prime}},{\bf\Omega^{\prime}},t).
\label{eq:torque}
\end{eqnarray}
 Combining the above equations, the final expression cross-correlation can be 
written as,
\begin{eqnarray}
<\Delta\omega_{VR}({\bf r},{\bf\Omega},t)\Delta\omega_{\rho}({\bf r},0)>=-A(k_{B}
T/\hbar)\Large[\int d{\bf r^{\prime}}c({\bf r}-{\bf r^{\prime}})\left<J^{2}({\bf r},
{\bf\Omega},0)\delta\rho({\bf r^{\prime}},0)\right>\nonumber \\+ 2\int dt^{\prime}
\int d{\bf r^{\prime}}\int d{\bf r^{\prime\prime}}d{\bf\Omega^{\prime}}c({\bf r}-
{\bf r^{\prime}}){\bf\nabla_{\Omega^{\prime}}}\cdot c({\bf r}-{\bf r^{\prime
\prime}},{\bf\Omega},{\bf\Omega^{\prime}})\left<{\bf J}({\bf r},{\bf\Omega},0)\delta
\rho({\bf r^{\prime\prime}},{\bf\Omega^{\prime}},t^{\prime})\delta\rho({\bf 
r^{\prime}},0)\right>\nonumber \\+ \int dt^{\prime}\int dt^{\prime\prime}\int d{\bf 
r^{\prime\prime}}d{\bf\Omega^{\prime}}\int d{\bf r^{\prime\prime\prime}}d{\bf
\Omega^{\prime\prime}}\int d{\bf r^{\prime}}c({\bf r}-{\bf r^{\prime}}){\bf
\nabla_{\Omega^{\prime}}}\cdot{\bf\nabla_{\Omega^{\prime\prime}}}c({\bf r}-{\bf 
r^{\prime\prime}},{\bf\Omega},{\bf\Omega^{\prime}})c({\bf r}-{\bf r^{\prime\prime
\prime}},{\bf\Omega},{\bf\Omega^{\prime\prime}})\nonumber\\\times\left<\delta\rho({
\bf r^{\prime\prime}},{\bf\Omega^{\prime}},t^{\prime})\delta\rho({\bf r^{\prime
\prime\prime}},{\bf\Omega^{\prime\prime}},t^{\prime\prime})\delta\rho(
{\bf r^{\prime}},0)\right>\Large]\nonumber\\\equiv{\bf I+II+III}.
\label{eq:final}
\end{eqnarray}
\end{widetext}

  We now analyze Eq.~(\ref{eq:final}) term by term. Since both $J$ and 
$\delta\rho$ are uncoupled at the same time, the contribution of {\bf I} 
is zero. The second term {\bf II} in Eq.~(\ref{eq:final}) consists of two 
density terms with an angular momentum term. Using cumulant expansion we 
can show that the contribution of {\bf II} is also equal to zero.

 To evaluate the third term ({\bf III}), we factorize the translational and 
angular variables in the two-particle direct correlation function (DCF) and we 
can write
\begin{eqnarray}
c({\bf r}-{\bf r^{\prime}},{\bf\Omega},{\bf\Omega^{\prime}}) = c_{\theta}
({\bf\Omega},{\bf\Omega^{\prime}})c_{0}(|{\bf r}-{\bf r^{\prime}}|).
\end{eqnarray}
Where $c_{0}(|{\bf r}-{\bf r^{\prime}}|)$ is the isotropic part of the direct 
correlation function. When the temperature $T\rightarrow T_{c}$, $c_{0}
\rightarrow\rho_{0}^{-1}$, which is finite. The angular part of the direct 
correlation function $c_{\theta}({\bf\Omega},{\bf\Omega^{\prime}})$ can be 
expanded in terms of the spherical harmonics.
\begin{eqnarray}
c_{\theta}({\bf\Omega},{\bf\Omega^{\prime}}) = \sum_{l_{1},l_{2},m}a_{l_{1}
l_{2}m}{\bf Y}_{l_{1},m}({\bf\Omega}){\bf Y}_{l_{2},m}({\bf\Omega^{\prime}}).
\end{eqnarray}
If we assume that the density effects are included in the spatial part $c_{0}(|
{\bf r}-{\bf r^{\prime}}|)$ then the final expression of the angular part of the 
direct correlation function will be finite and nonzero. If one assumes that
the $\delta\rho$ are Gaussian random variables, then the value of the triplet 
correlation function in {\bf III} is zero, by cumulant expansion. However, the 
distribution of the fluctuation in number density clearly shows a non-Gaussian 
behavior, arising from the long lived heterogeneity near critical point 
(see Fig.~$11$). We, therefore, expect a nonzero value of the 
cross-correlation between VR coupling and atom-atom terms from the triplet 
correlation function in Eq.~(\ref{eq:final}). Moreover, the amplitude of the
cross-correlation is predicted to be small at all temperatures away from $T_{c}$
but it is predicted to become significant as $T_{c}$ is approached, where
the density fluctuation becomes non-Gaussian.

  We have calculated the term $\Delta R$ in Eq.~\ref{eq:omegavr2} (see Appendix B)
and found it to be always positive. It is obvious that $J^{2}>0$. So we can 
conclude that $\delta\omega_{VR}>0$ {\large[see Fig.~$12b$\large]}. 
Figure~$12(a)$ shows that most of the distribution of fluctuation in 
$\delta\omega_{\rho}$ are in the negative direction. On the other hand the 
distribution of fluctuation in $\delta\omega_{VR}$ is in the positive direction 
{\large[see Fig.~$12b$\large]}. Here, $\Delta\omega_{\rho} = \left({\it n}
\hbar(-{\it f})/2\mu^{2}\omega^{3}_{0}\right){\it F}^{i}_{1Q} + \left({
\it n}\hbar/{2\mu\omega_{0}}\right){\it F}^{i}_{2Q}.$ So the distribution of 
cross-correlation between density and VR coupling terms must be negative (see 
Fig.~$13$). 

  The above analysis  MCT demonstrates that the large enhancement of 
vibration-rotation coupling near the gas-liquid critical point arises from 
the non-Gaussian behavior of density fluctuation and this enters through a 
nonzero value of the triplet direct correlation function {\large[Eq.~(\ref{eq:final})\large]}.

\section{Divergence of Raman linewidth near the CP}

  Along the coexistence line and the critical isochore, we have found that the 
temperature dependence of the linewidth ($\Gamma$) is singular near the critical 
point. The divergence-like rise near $T_{c}$ is fitted to a form $(T-T_{c})^{-
\beta^{\prime}}$. From the fitting with both experimental and theoretical data 
we have found $\beta^{\prime} = 0.386$ along the coexistence line and $\beta^{
\prime} = 0.207$ along the critical isochore.  

  Above $T_{c}$, the fluctuations \cite{swapan} are small which cause the resonance 
lines to be homogeneously broadened and the line shape is Lorentzian. On 
approaching $T_{c}$, the fluctuations \cite{md1,md2} become large and their 
correlation time also increases, as a consequence the line shape becomes Gaussian.

  The order parameter for the liquid-gas critical point is $(\rho_{L}-\rho_{G})$.
This goes to zero as $(T_{c}-T)^{-\beta}$ where $\beta\simeq\frac{1}{3}$. Mukamel
{\it et al.} \cite{muk} have studied the broadening of spectral lines by using 
mode-coupling theory \cite{ks,fixman} near a liquid-gas critical point. They 
assumed Lorentzian form of the line and found that the linewidth increases with 
$\epsilon^{-s}$, where $\epsilon = |(T -T_{c})/T_{c}|$. They have also found 
that the value of critical exponent (s) changes from 0.607 for large $\epsilon$ 
to $0$ for small $\epsilon$. That is, the singularity becomes weaker as $T_{c}$ 
is approached. It is likely that simulations are not sufficiently close to 
$T_{c}$ because simulations are finite sized.

         The reason for the relative success of a small system in reproducing 
experimental behavior is that dephasing probes only local static fluctuations.
Thus long range static and dynamic correlations important in specific heat, 
compressibility, or light scattering are not relevant in vibrational dephasing. 
Second, while our simulated system can certainly capture the increase in 
fluctuation, it can never capture the long range fluctuations. But it captures 
enough to reproduce many of the features. Note that even a small system is 
capable of exhibiting large fluctuations near the gas-liquid coexistence/critical 
point. Put bluntly, truly large scale critical fluctuations are neither 
required nor reflected for dephasing simply because dephasing is a local 
process.

\section{Conclusion}

    In this article extensive MD simulations of vibrational phase relaxation of 
nitrogen have been presented. The simulations reported here seem to reproduce 
many of the anomalies observed in experiments. It shows (a) the origin of 
enhanced negative cross-correlations, (b) a crossover from a Lorentzian-type
to a Gaussian line shape as the critical point is approached, (c) the non-monotonic 
dependence of rms frequency fluctuation on temperature along the critical 
isochore, (d) a lambda shaped dependence of linewidth on temperature, and (e) a 
near divergence of linewidth near the critical point.

\section{Acknowledgment}

  One of us (S.R.) would like to thank A. Mukherjee, Prasanth. P. Jose, and 
R. K. Murarka for helpful discussions. S.R. acknowledges the CSIR (India)
for financial support. This work was supported in part by grants from CSIR, 
India and DAE, India.

\section{Appendix A: Derivative of Lennard-Jones potential}

  The derivative of the LJ potential can be calculated as 

\begin{widetext}
\begin{eqnarray}
\frac{\partial  v_{ij}}{\partial q_{i}}&=&4\epsilon\sum_{\alpha,\beta = 1}^{2}
\Large[\Large\{1+\gamma(1+\delta_{ij}) + 2\gamma^{2}(q_{j} + q_{i}\delta_{ij})\}\times
\{\left(\frac{\sigma}{r_{i\alpha j\beta}}\right)^{12}(1+\delta q_{i}+\delta 
q_{j})^{12}\nonumber\\&-&\left(\frac{\sigma}{r_{i\alpha j\beta}}\right)^{6}
(1+\delta q_{i}+ \delta q_{j})^{6}\}+\{1+ \gamma(q_{i}+q_{j})+2\gamma^{2}q_{i}
q_{j}\}\nonumber\\&\times&\{\frac{d}{dr_{i\alpha j\beta}}\left(\frac{\sigma}
{r_{i\alpha j\beta}}\right)^{12}\frac{\partial r_{i\alpha j\beta}}{\partial q_{i}}
(1+\delta q_{i}+\delta q_{j})^{12}\nonumber\\&+&\left(\frac{\sigma}{r_{i\alpha 
j\beta}}\right)^{12}(1+\delta+\delta\delta _{ij})\times 12(1+\delta q_{i} + 
\delta q_{j})^{11}\nonumber\\&-&\frac{d}{dr_{i\alpha j\beta}}\left(\frac{\sigma}
{r_{i\alpha j\beta}}\right)^{6}\frac{\partial r_{i\alpha j\beta}}{\partial q_{i}}
(1 + \delta +\delta q_{i}+\delta q_{j})^{6}\nonumber\\&-&\left(\frac{\sigma}
{r_{i\alpha j\beta}}\right)^{6}\times6(1+\delta+\delta\delta_{ij})(1+ \delta 
+\delta q_{i}+\delta q_{j})^{5}\Large\}\Large],
\end{eqnarray}
\end{widetext}
[where 
$\frac{d}{dr_{i\alpha j\beta}}\left(\frac{\sigma}{r_{i\alpha j\beta}}\right)^{12}
= -\frac{12}{\sigma}\left(\frac{\sigma}{r_{i\alpha j\beta}}\right)^{13} $
and
$\frac{d}{dr_{i\alpha j\beta}}\left(\frac{\sigma}{r_{i\alpha j\beta}}\right)^{6}
= -\frac{6}{\sigma}\left(\frac{\sigma}{r_{i\alpha j\beta}}\right)^{7}]$.

\begin{widetext}
\begin {eqnarray}
\frac{\partial v_{ij}}{\partial q_{j}}&=&4\epsilon\sum_{\alpha,\beta = 1}^{2}
\Large[\Large\{1+\gamma(1+\delta_{ij}) + 2\gamma^{2}(q_{i} + q_{j}\delta_{ij})\}\times\{
\left(\frac{\sigma}{r_{i\alpha j\beta}}\right)^{12}(1+\delta q_{i}+\delta q_{j})
^{12}\nonumber\\&-&\left(\frac{\sigma}{r_{i\alpha j\beta}}\right)^{6}(1+\delta 
q_{i}+ \delta q_{j})^{6}\}+ \{1+ \gamma(q_{i}+q_{j})+2 \gamma^{2}q_{i}q_{j}\}
\nonumber\\&\times&\{\frac{d}{dr_{i\alpha j\beta}}\left(\frac{\sigma}{r_{i\alpha j\beta}}
\right)^{12}\frac{\partial r_{i\alpha j\beta}}{\partial q_{j}}(1+\delta q_{i}+
\delta q_{j})^{12}\nonumber\\&+&\left(\frac{\sigma}{r_{i\alpha j\beta}}\right)^
{12}(1+\delta+\delta\delta_{ij})\times12(1+\delta q_{i} + \delta q_{j})^{11}
\nonumber\\&-&\frac{d}{dr_{i\alpha j\beta}}\left(\frac{\sigma}{r_{i\alpha j\beta}}
\right)^{6}\frac{\partial r_{i\alpha j\beta}}{\partial q_{j}}(1 + \delta+\delta 
q_{i}+\delta q_{j})^{6}\nonumber\\&-&\left(\frac{\sigma}{r_{i\alpha j\beta}}
\right)^{6}\times6(1+\delta+\delta\delta_{ij})\times(1+\delta +\delta q_{i}+
\delta q_{j})^{5}\Large\}\Large].
\end{eqnarray}
\end{widetext}

\section{Appendix B: Frequency modulation from
vibration-rotation coupling}

  The vibration-rotation centrifugal coupling term is $E_{vib-rot}=
\frac{J^{2}}{2I}$, where $I(=2\mu r^{2})$ is the moment of inertia and 
$\mu$ is the reduced mass of a diatomic molecule. This VR term is 
important only when $I$ depends on vibrational coordinate $(q)$ through 
the bond length, i.e., $r(q) = r_{0}+\Delta r(q)$. Expanding this as a 
Taylor series about the equilibrium bond length $r_{0}$, the $E_{vib-rot}$
can be rewritten as
\begin{eqnarray}
E_{vib-rot}&=&\frac{J^{2}}{2\mu(r_{0}+\Delta r)^2}\nonumber \\
&=&\frac{J^{2}}{2\mu r_{0}^{2}}\left(1+\frac{\Delta r}{r_{0}}\right)^{-2}
\nonumber \\
&=&\frac{J^{2}}{2I_{0}}\left[1-\frac{2\Delta r}{r_{0}}+3\left(\frac{\Delta r}
{r_{0}}\right)^{2}-\ldots \right]. 
\end{eqnarray}

 The contribution of VR coupling to the total $C_{\omega}(t)$ is given by
$C^{VR}_{\omega}$ as
\begin{eqnarray}
C^{VR}_{\omega}(t) = \left<\Delta\omega_{VR}(t)\Delta\omega_{VR}(0)\right>
\end{eqnarray}
where
\begin{eqnarray}
\Delta\omega_{VR}(t)&=&\left[E_{11}(t)-E_{00}(t)\right]/\hbar\nonumber \\
&=&\left(\frac{\Delta R}{\hbar I_{0}r_{0}}\right)\Delta J^{2}(t)
\label{eq:omegavr2},
\end{eqnarray}
where $\Delta R = \left(Q_{11}-Q_{00}\right) - \frac{3\left(Q^{2}_{11}-Q^{2}_{00}
\right)}{2r_{0}}$, $Q_{nn}=<n|\Delta r|n>$ and  $Q^{2}_{nn}=\left<n|(\Delta r)^{2}|n\right>$
are the expectation values of bond length displacement and its square and 
$\Delta J^{2}(t) = J^{2}(t) - \left<J^{2}(0)\right>$. We used $\left<J^{2}\right> = 2I_{0}k_{B}T$.

\end{document}